%% file: main.tex
\documentclass[nonacm, manuscript]{acmart}

\AtBeginDocument{%
  \providecommand\BibTeX{{%
    \normalfont B\kern-0.5em{\scshape i\kern-0.25em b}\kern-0.8em\TeX}}}

\usepackage{graphicx}
\usepackage{multirow}
\usepackage{array}
\usepackage{rotating}
\usepackage{multirow,tabularx}
\usepackage{rotating}  
\usepackage{lscape}
\usepackage{xcolor}
\usepackage{booktabs} 
\usepackage{subcaption}

\usepackage{longtable}
\usepackage{ltablex} 
\usepackage{tabularx}
\usepackage{ragged2e}
\usepackage{caption}
\usepackage{siunitx}
\newcolumntype{Y}{>{\RaggedRight\arraybackslash}X} 
\newcolumntype{L}[1]{>{\raggedright\arraybackslash}p{#1}}
\newcolumntype{C}[1]{>{\centering\arraybackslash}p{#1}}
\usepackage[most]{tcolorbox}
\usepackage{enumitem}

\newcommand{\parisa}[1]{{\small\textcolor{blue}{\bf [* Parisa: #1]}}}

\newtcolorbox{callout}{
  enhanced, breakable,
  colback=black!2, colframe=black!50,
  boxrule=0.6pt, arc=8pt,
  left=8pt,right=8pt,top=6pt,bottom=6pt
}

\begin{document}

\title[The Social Gaze of LLMs]{The Social Gaze of LLMs: A Literature Review of Multimodal Approaches to Human Behavior Understanding}

\author{Zihan Liu}
\affiliation{%
  \institution{University of Illinois Urbana-Champaign}
  \country{USA}}

  \author{Parisa Rabbani}
\affiliation{%
  \institution{University of Illinois Urbana-Champaign}
  \country{USA}}

    \author{Veda Duddu}
\affiliation{%
  \institution{University of Illinois Urbana-Champaign}
  \country{USA}}
  
      \author{Kyle Fan}
\affiliation{%
  \institution{University of Illinois Urbana-Champaign}
  \country{USA}}

  \author{Madison Lee}
\affiliation{%
  \institution{University of Illinois Urbana-Champaign}
  \country{USA}}

    \author{Yun Huang}
\affiliation{%
  \institution{University of Illinois Urbana-Champaign}
  \country{USA}}

\begin{abstract}

\input{section/0Abstract}

\end{abstract}




\maketitle

\section{Introduction}
\input{section/1Introduction}

\section{Related Work}
\input{section/2relatedworks}

\section{Method}
\input{section/3Method}

\input{section/4Results-rq1}
\input{section/4Results-rq2}
\input{section/4Results-rq3}
\input{section/4Results-rq4}

\section{Discussion}
In this section, we explore the research gaps identified across four key perspectives: the application of LLM-based social intelligence, technological operations, system evaluation, and ethical considerations. Building on these observations, we propose a forward-looking research agenda for each area individually, as well as for the broader field as a whole.

\input{section/5Discussion}

\input{section/5Discussion-whole}

\bibliographystyle{ACM-Reference-Format}
\bibliography{chi}

\end{document}

%% file: section/0Abstract.tex
LLM-powered multimodal systems are increasingly used to interpret human behavior, yet how researchers apply the models' "social competence" remains poorly understood. This paper presents a systematic literature review of 176 publications across different application domains (e.g., healthcare, education, and entertainment). Using a four-dimensional coding framework (application, technical, evaluative, and ethical), we find (1) frequent use of pattern recognition and information extraction from multimodal sources, but limited support for adaptive, interactive reasoning; (2) a dominant "modality-to-text" pipeline that privileges language over rich audiovisual cues, striping away nuanced social cues; (3) evaluation practices reliant on static benchmarks, with socially grounded, human-centered assessments rare; and (4) Ethical discussions focused mainly on legal and rights-related risks (e.g., privacy), leaving societal risks (e.g., deception) overlooked—or at best acknowledged but left unaddressed. We outline a research agenda for evaluating socially competent, ethically informed, and interaction-aware multi-modal systems.

%% file: section/1Introduction.tex
Human social interaction is inherently multimodal, unfolding through complex combinations of verbal utterances, facial expressions, gestures, gaze patterns, and vocal prosody \cite{lee2024towards}. Understanding these rich social signals is not just a perceptual challenge; instead, it lies at the heart of what makes humans socially intelligent. Social intelligence, broadly defined as the ability to perceive, interpret, and act appropriately in social contexts \cite{thorndike1920intelligence, vernon1933some}, is a foundational dimension of human intelligence and a critical capability for AI systems that aim to interact meaningfully with people \cite{fiore2013toward, wiltshire2014interdisciplinary, lee2024towards}.

The emergence of AI systems that integrate Large Language Models (LLMs) with multimodal perception capabilities marks a significant shift in how machines can understand human behavior. Unlike traditional approaches that rely on isolated modality-specific model \cite{li2025perception}, LLM-powered multimodal systems  leverage the rich semantic understanding and reasoning abilities that LLMs have developed through large-scale language training. Drawing an analogy to human perception, modality modules such as visual or audio encoders act like eyes and ears, capturing and pre-processing visual and acoustic signals, while the LLM serves as the brain, integrating these inputs and reasoning over their social meaning \cite{Yin_2024}. For example, LLM-powered multimodal systems are being used to analyze student engagement patterns in educational settings  \cite{yu2024raw}, assess social development in children with autism spectrum disorders \cite{deng2024hear}, and summarize collaborative strategies in team sports \cite{lee2024sportify}.

However, little is known about how researchers have applied the models’ social competence in designing these systems. 
Several challenges limit our understandings of the research landscape. First, multiple research traditions approach this challenge from diverging perspectives: psychologists emphasize multidimensional constructs \cite{conzelmann2013new, weis2005social}, AI researchers build technical systems around isolated cues \cite{muller2024recognizing}, and evaluation remains dominated by static benchmarks \cite{agrawal2024listen}. Because of this fragmentation, there is no coherent framework for AI system designers, developers, and evaluators to understand what  “social intelligence” means for AI in multimodal systems, how it is technically realized, and how it should be assessed. Moreover, a disconnection persists between technical and applied communities: AI research emphasizes benchmark progress, while fields such as HCI, education, and healthcare lack clear frameworks for interpreting what these systems can and cannot do in socially meaningful contexts. These gaps underscore the need for a systematic review that bridges disciplinary boundaries and provides a structured account of how LLM-powered multimodal systems engage with human social behavior.

This paper addresses the  gap through a systematic literature review of LLM-powered multimodal systems that are designed to understand human behavior. 
Our analysis synthesizes research across technical venues in AI and applied communities such as HCI, education, and healthcare. 
Specifically, we examine the following four perspectives: 
\begin{itemize}
    \item \textbf{RQ1} (application): \textit{What is the application scope of LLM-powered multimodal systems for human behavior understanding in terms of target behaviors, interaction contexts, and operationalized social competencies?}
    \item \textbf{RQ2} (technical): \textit{How are LLM-powered multimodal systems technically operationalized for human behavior understanding?}
    \item \textbf{RQ3} (evaluative): \textit{What methodologies and metrics are employed to evaluate the performance of LLM-powered multimodal systems in understanding human behavior?}
    \item \textbf{RQ4} (ethical): \textit{What are the primary ethical challenges and risks associated with developing and deploying LLM-powered multimodal systems for human behavior understanding?}
\end{itemize}

This paper makes a timely contribution to the HCI community by presenting a thorough review of LLM-enabled social intelligence applications. We synthesize, systematize, and critically assess this fractured landscape, yielding important implications for design, technical, and applied communities:
\begin{itemize}
    \item \textbf{A Systematic, Interdisciplinary Synthesis of 176 Studies.} We provide the first large-scale literature review of LLM-powered multimodal systems for human social behavior understanding, bridging fragmented work across AI, HCI, education, and healthcare to reveal patterns, gaps, and opportunities. 

  \item \textbf{A Four-Dimensional Coding Framework for Social Intelligence:} 
    We introduce and apply a four-dimensional coding framework---covering \emph{application}, \emph{technical}, \emph{evaluative}, and \emph{ethical} perspectives---to systematically characterize how social intelligence is operationalized in LLM-powered multimodal systems. This framework offers researchers a structured vocabulary to guide future system design, benchmarking, and governance.

    \item \textbf{Empirical Insights into the Current Landscape:} 
    Our analysis reveals several key takeaways: (1) Nearly all these research focus on social perception (100\%) and reasoning (95\%), while social interaction (41.8\%) and creativity (18\%) are rarely implemented; (2) a \emph{modality-to-text bottleneck} that compresses rich multimodal signals into transcripts or keyframes before reasoning, losing rich multimodal cues; (3) evaluation practices that remain \emph{machine-centric}, with relatively few longitudinal, interactive, or human-centered studies; and (4) 45\% of reviewed papers do not mention ethics and risks. 

    \item \textbf{Research Agendas for Socially Competent and Ethical LLM-powered Multimodal Systems:} 
    Building on these findings, we chart a research agenda emphasizing (1) richer multimodal integration that preserves more modality information--prosody and gaze, etc; (2) development of socially grounded, human-centered, and longitudinal evaluation protocols; (3) diversification of model families and adoption of hybrid operationalization strategies; and (4) systematic mitigation of fairness, bias, and misuse risks beyond privacy, enabling accountable and socially situated AI behavior. We also consolidate a directory of benchmarks and datasets (Table \ref{tab:bench_mark_table}), mapping human goals to computational tasks and evaluation resources, which can be used to evaluate LLM-powered multimodal systems comprehensively.
\end{itemize}

%% file: section/2relatedworks.tex
\subsection{Social Intelligence and Social AI} 

The concept of social intelligence was first introduced by Thorndike (1920) as the ability to “understand and manage men and women, boys and girls, and to act wisely in human relations” \cite{thorndike1920intelligence}. Vernon defined social intelligence as "knowledge of social matters and insight into the moods or personality traits of strangers" and as the ability to "get along with others and easein society" \cite{vernon1933some}. Importantly, both of these two definition highlighted a dual nature: a cognitive facet (understanding others) and a behavioral facet (acting effectively in social situations). Building on this, subsequent definitions alternated between the two aspects \cite{weis2005social, hendricks1969measuring}. For instance, Wedeck \cite{wedeck1947relationship} conceptualizes social intelligence from a cognitive perspective, defining it as “the capacity to judge correctly the feelings, moods, and motivations of individuals.” 

Building on these psychological foundations, researchers have increasingly asked how such social competencies might be implemented in computational systems, giving rise to the field of social AI or artificial social intelligence (ASI). Social AI is often defined as the pursuit of “socially-intelligent AI agents” \cite{lee2024towards}, while ASI is framed as the capability to help these agents calibrate the outputs of their internal models to be understood by humans \cite{xu2024academically}. Scholars identify three key capabilities necessary for effective social understanding: the ability to interpret multimodal social cues, to reason about multi-party dynamics, and to infer beliefs or mental states \cite{lee2024towards}. Mathur et al. (2024) further argues that building socially intelligent AI requires agents that can not only sense and reason about human affect, behavior, and cognition, but also adapt dynamically across diverse social contexts  \cite{mathur2024advancing}. Efforts to engineer ASI have thus focused on how social information is embodied in behavior. Researchers stress that an agent must grasp that intentions, emotions, and personalities are expressed through verbal and non-verbal cues, and therefore require integrated processing pipelines that link perception with social reasoning \cite{williams2022supporting, fiore2013toward, wiltshire2014interdisciplinary}. These discussions connect directly to broader questions of alignment and governance: just as individuals with high social intelligence can manage conflicts between personal and group objectives and avoid toxic behaviors, socially intelligent AI is envisioned as a path toward systems that are both norm-sensitive and collaborative \cite{korinek2022aligned, albrecht2009social}. 

Despite these advances, the notion of social intelligence in AI remains fragmented and inconsistently defined. Psychology has long emphasized social intelligence as a multidimensional construct encompassing perception, memory, reasoning, and behavior \cite{conzelmann2013new}. However, AI research often narrows this scope to isolated elements such as theory of mind \cite{williams2022supporting} or social cue recognition \cite{lee2024towards}, without integrating them into a coherent framework. This conceptual gap motivates our review: to clarify how social intelligence has been framed in AI research and to prepare the ground for analyzing how systems have evolved from unimodal to multimodal approaches.

\subsection{Social AI: from Uni-modal to Multimodal }

Early efforts in social AI were largely unimodal \cite{tomarUnimodal} and primarily focused on descriptive recognition of isolated social cues \cite{vinciarelli2008social}. Natural language processing systems classified utterances as polite or impolite, empathetic or neutral \cite{sap2022neural}. Computer vision models identified emotions from facial expressions or categorized gestures. Speech analysis models mapped vocal prosody to affective states \cite{feng2024foundation}. These systems demonstrated that isolated social cues could be computationally detected, but reduced social intelligence to recognition tasks, offering little capacity to interpret meaning or context.

The development of LLMs created a significant shift in AI social capabilities bringing explicit reasoning and interactive abilities \cite{berti2025emergentabilitieslargelanguage}. Unlike unimodal classifiers that simply detected the presence of signals, LLM-based systems could leverage prior knowledge and existing instructions to draw inferences about moral reasoning, intent detection, and social commonsense \cite{takemoto2024moral, arora2024intentdetectionagellms, mousavi2025garbageinreasoningout}. From this perspective, LLMs enabled a move from answering the question "what signal is present?" to asking "why does this social signal matter in this context?", marking an important step toward artificial social intelligence \cite{li2025perception}. At the same time, their reasoning remains fundamentally constrained: because these models operate only over textual descriptions, they may fail to capture the implicit cues, sarcasm, and non-verbal signals that are central to human social understanding \cite{chang2025bridginggapllmshuman, castro2019towards}.

Overcoming the limitations of both classic unimodal models and text-only LLMs, recent research has turned toward multimodal AI systems that incorporate LLMs, aiming to integrate textual, visual, and auditory channels for more socially grounded understanding \cite{lee2024towards, huang2024surveyevaluationmultimodallarge}. These systems utilize LLMs as a reasoning brain and encode other data types, allowing the models to perceive and reason about the world while capturing the same information that humans do through senses \cite{Yin_2024}. As Li et al. observe, this reflects a broader paradigm shift in multimodal reasoning: from perception-driven modular systems, where reasoning was implicit within task-specific classifiers, to language-centric reasoning frameworks, where inference is articulated through structured prompts and extended chains of thought \cite{li2025perception}.

The trajectory from unimodal recognition to LLM-enabled reasoning to multimodal integration illustrates the rapid evolution of social AI. Yet the technical landscape remains fragmented: systems are often built as loosely coupled pipelines \cite{li2025perception}, with language reasoning dominating \cite{mathur2024advancing}, and integration across modalities limited in depth \cite{li2024multimodal}. These constraints highlight a technical gap between the richness of human social interaction and the limited architectures  used to approximate it, setting the stage for our subsequent discussion of evaluation. 

\subsection{Evaluating Social Intelligence in AI Systems}
\label{rw:eval}
The rise of multimodal AI systems that incorporate LLMs has been accompanied by numerous surveys cataloging their architectures, training techniques, and performance on general-purpose benchmarks \cite{liang2024comprehensive, yin2024survey, li2024multimodal}. However, existing surveys are often general or technically focused, lacking attention to social intelligence. When included in such surveys, social intelligence is often not properly evaluated to capture the dynamic, interactive, and context-dependent nature of real social situations. Without robust, socially grounded evaluation frameworks, we cannot truly measure the capabilities or limitations of these systems in understanding human social dynamics \cite{ge2024mllmbenchevaluatingmultimodalllms}. We risk developing models that perform well on simplified benchmarks but fail in real-world social interactions. 

Current evaluation practices in the AI community have largely followed a technical benchmark tradition, breaking social intelligence down into focused, task-specific capabilities. Tasks such as emotion recognition or intent prediction are typically framed as classification problems, with performance reported as a single accuracy score \cite{chakhtouna2024multi, li2023intentqa}. Early benchmarks like Social-IQ \cite{zadeh2019social} provided models with video clips and multiple-choice questions, enabling quantitative comparisons but also allowing models to exploit dataset biases without engaging in genuine social understanding. More recent efforts have tried to address these shortcomings. SIV-Bench \cite{kong2025siv}, for example, decomposes the evaluation into three components: Social Scene Understanding (SSU), Social State Reasoning (SSR), and Social Dynamics Prediction (SDP). SocialMaze \cite{xu2025socialmaze} goes further by introducing multi-turn, game-like interaction scenarios such as hidden role deduction, aiming to capture more strategic forms of social reasoning. While these benchmarks represent progress, they remain limited to structured task outcomes and continue to overlook the reciprocal, adaptive, and situated qualities that characterize real social intelligence \cite{blili2024unsocial, zhou2025socialeval}.

As Mathur et al. (2024) emphasize, evaluating artificial social intelligence involves challenges that go beyond technical benchmarking: the ambiguity of defining social constructs, the subtlety of multimodal signals, the need to represent multiple perspectives, and the requirement of agency and adaptation across contexts \cite{mathur2024advancing}. Current approaches fall short in each of these dimensions. This uneven landscape motivates our review: to synthesize how existing studies evaluate social intelligence in multimodal AI systems, to identify where evaluations succeed, and to highlight where socially grounded approaches are still lacking.

%% file: section/3Method.tex
This literature review aims to investigate the emerging capabilities, limitations, and evaluation practices associated with social intelligence in LLM-powered multimodal systems. We specifically focused on synthesizing research that explores how these systems perceive, process, and interpret human social interactions through externally observable, non-verbal social cues (e.g. visual and auditory signals). 

Based on this focus, we defined the following inclusion criteria. (a) \textit{Model Architecture}: The paper must describe a system or method that integrates a LLM component as part of its architecture. 
  (b) \textit{Multimodal social cues}: This system must be designed to process or analyze non-verbal social cues (e.g., gaze, facial expression, prosody, body posture) from vision or audio modalities in the context of understanding or engaging in human social interactions. 
 (c) \textit{Contribution Type}: We include papers that make a relevant contribution to the field of socially intelligent multimodal models.

We excluded papers that met any of the following categories: (a) \textit{Text-Only Focus}: Works that relied solely on textual input without addressing nonverbal cues were excluded.
  (b) \textit{Biosensor Data Emphasis}: We excluded studies that primarily rely on biosensor data—such as heart rate or skin conductance—captured via wearable devices (e.g., \cite{schmidt2018introducing}). While such data is relevant in affective computing, it is not directly observable in typical face-to-face human interaction. Our review focuses on externally perceivable social cues (e.g., facial expressions, gestures, vocal tone) that can be interpreted through visual and auditory modalities in natural social settings. 

\begin{figure}[h]
    \centering
    \includegraphics[width=.95\linewidth]{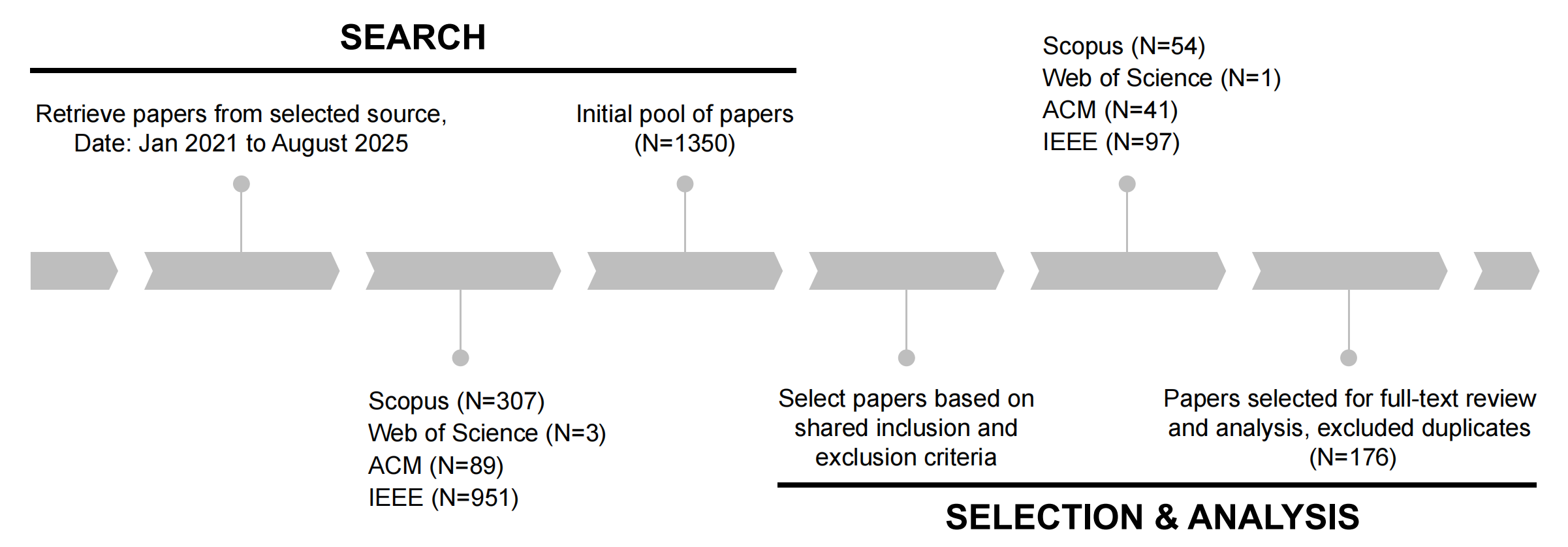}
    \caption{\textbf{Workflow of the Literature Search and Selection Process.} This flowchart illustrates the four-stage methodology used to identify relevant papers. The process began with an initial pool of 1,350 papers from four databases, which was narrowed down based on inclusion and exclusion criteria, ultimately resulting in a final corpus of 176 papers for analysis.}
    \label{fig:workflow diagram}
\end{figure}

Eventually, the eligible contributions encompass a range of types, including: the introduction of a new model or algorithm; the development of a system, application, or toolkit; the proposal of a conceptual framework, architecture, or processing pipeline; or the execution of an empirical study that evaluates or applies LLM-powered multimodal AI system in social contexts.
Our methodology follows a three-stage process—Search, Selection, and Analysis—which is visually summarized in Figure \ref{fig:workflow diagram}. The subsequent subsections detail each of these stages. 

\subsection{Search}
Our search strategy was structured around three main categories of keywords: (1) technical model terms, (2) social behavior terms, and (3) media modality terms. For the technical terms, we included variations such as: \textit{"large language model"},  \textit{"vision language model"}, \textit{"visual language model"}, and \textit{"large multimodal model"}. To capture work related to social intelligence, we used keywords describing social behavior, including: \textit{"social interaction"}, \textit{"social behavior"}, \textit{"social cue"}, \textit{"social signal"}, \textit{"human interaction"}, and \textit{"human behavior"}. Finally, we added media modality terms to narrow the scope to video-based analysis, using phrases such as: \textit{"video understanding"}, \textit{"audio understanding"}, \textit{"audio analysis"}, \textit{"voice analysis"}, \textit{"voice understanding"}, \textit{"speech analysis"}, and \textit{"speech understanding"}. These terms were combined using Boolean logic to construct flexible search queries, which were tailored to the syntax and capabilities of each database.

We searched four databases to ensure comprehensive coverage: the Association for Computing Machinery Digital Library (ACM DL), Scopus, Web of Science, and IEEE. The initial search covered papers published from January 2021 to August 2025 and resulted in 89 papers from ACM DL,  307 from scopus, 3 from Web of Science, and 951 from IEEE. After removing duplicates, there were a total of 1350 papers in the initial pool. 

\subsection{Screening and Selection}
To identify relevant papers, we conducted a full-text manual review of each candidate based on the inclusion and exclusion criteria outlined earlier. Rather than relying solely on titles and abstracts, we carefully examined the full content of each paper to determine whether it involved the use of a LLM-powered multimodal system to interpret or respond to externally observable nonverbal social signals. Notably, many papers addressed aspects of social interaction without explicitly framing their work as achieving "social intelligence." These were still considered if the system demonstrated relevant competencies. This screening process resulted in a final selection of 176 papers for analysis. 

\subsection{Coding Procedure}

To scale our qualitative analysis while preserving rigor, we adopted a human-in-the-loop, LLM-assisted coding workflow, drawing on recent best practices for AI-supported qualitative research (e.g., \cite{delgado2025transforming, xiao2023supporting, mcginness2025highlighting}). Five authors collaboratively coded an initial set of 15 papers to develop a structured codebook across four focal dimensions: application, technical, evaluative, and ethical. This human-coded set served as the ground truth for subsequent AI-assisted analysis.

We iteratively designed few-shot prompting templates and used the Gemini 2.5 Pro model (temperature = 0) to code the remaining papers. The LLM was instructed to extract supporting textual quotes and providing justifications before synthesizing final codes. This two-step protocol was designed to reduce hallucinations and enhance traceability \cite{mcginness2025highlighting}. All LLM-generated outputs were manually reviewed and corrected by human coders by reading justifications and quotes.

To evaluate the reliability of our pipeline, we conducted an inter-rater reliability (IRR) analysis on a random sample of 30 papers per coding category. LLM-assigned codes were compared against independent human annotations. Results showed substantial agreement (e.g. in \textit{Ethical Risks}, the micro-averaged Cohen’s kappa was 0.717), supporting the validity and credibility of our approach. Additional details on the full pipeline, including prompting strategies, model configuration, and full per-category IRR results, are provided in Appendix.

%% file: section/4Results-rq1.tex
Our systematic review and analysis of 176 selected papers reveal the current landscape of socially intelligent LLM-powered multimodal systems. The findings are structured around our four research questions, covering the scope of applications (RQ1), the technical operationalization (RQ2), evaluation practices (RQ3), and the primary challenges and risks (RQ4).

\section{RQ1: Scope and Nature of LLM-powered Multimodal Systems in Human Behavior Understanding}
\label{results:rq1}

\begin{figure}[h]
    \centering
    \includegraphics[width=0.9\linewidth]{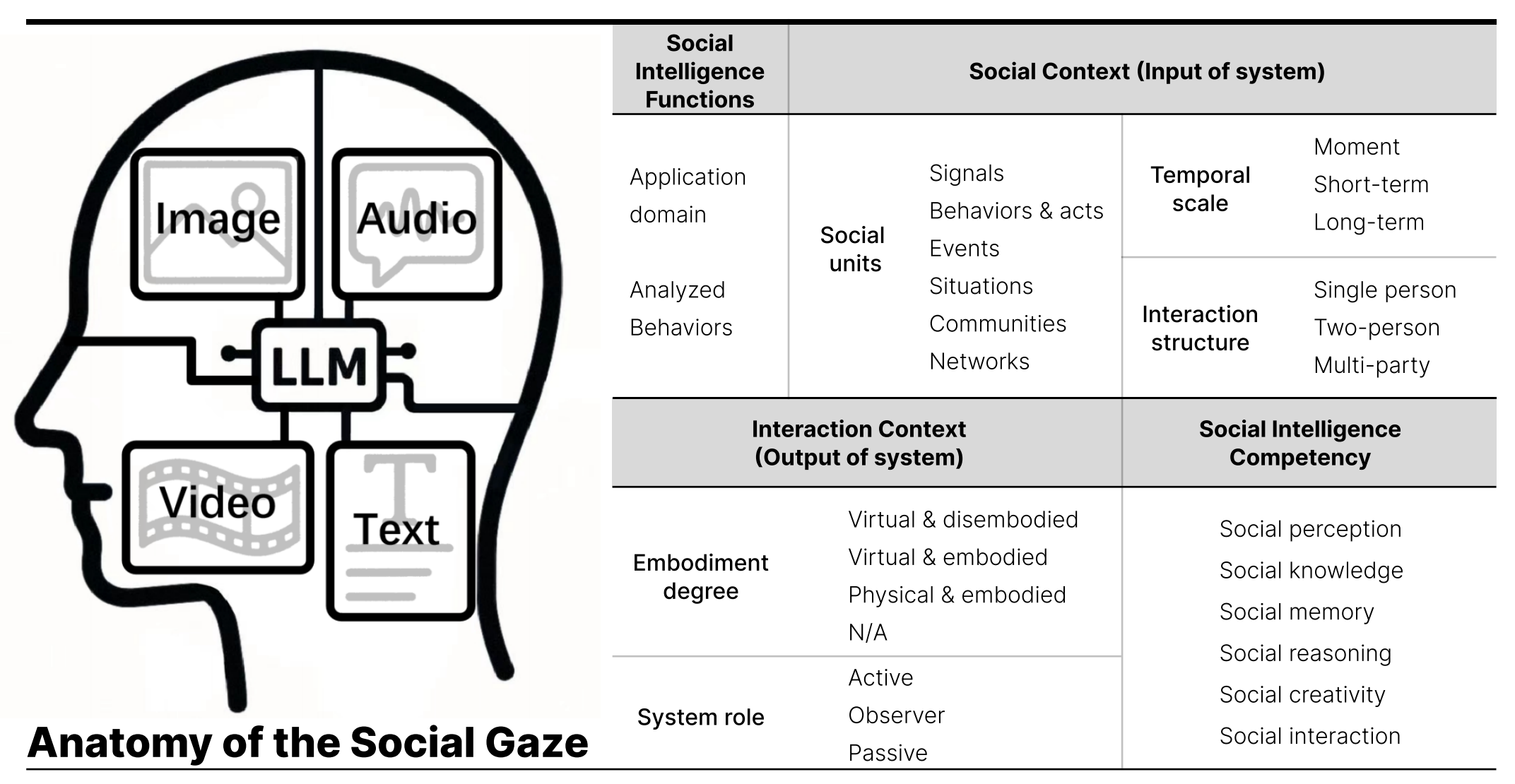}
    \caption{\textbf{A Conceptual Framework for Analyzing Social Intelligence (RQ1).} This diagram outlines the multi-dimensional coding framework used to analyze how social intelligence is applied in LLM-powered systems. It deconstructs the analysis into four key areas: the system's core Social Intelligence Functions, the Social Context it takes as input, the Interaction Context of its output, and the specific Social Intelligence Competencies it demonstrates.}
    \label{fig:anatomy}
\end{figure}

Our coding scheme for RQ1 was developed through a top-down process, beginning with theory-driven constructs derived from existing literature \cite{mathur2024advancing}, and iteratively refined through collaborative discussions among the research team. All coding categories for RQ1 are summarized in Figure \ref{fig:anatomy}.

\subsection{\textbf{Application Domains and Targeted Behavioral Cues}}
To map the landscape of LLM-powered multimodal systems in human behavior understanding, we analyzed both the application domains these systems target, and the types of human behaviors they are designed to recognize and interpret. 

In analyzing the application contexts of our reviewed systems, we found that existing research divides into two distinct but complementary threads: application-oriented research, which targets specific real-world application contexts, and research-oriented work, which focuses on advancing foundational knowledge or core technical capabilities without a direct application in mind. 

Among the application-oriented papers, the most prominent domain is Healthcare and Well-being, accounting for nearly a quarter of the corpus (23.73\%). These studies primarily target clinical or therapeutic settings, such as creating AI systems that reduce anxiety for autistic adolescents \cite{xie2024empathetic}. This is followed by applications in key societal sectors like Media, Entertainment, \& Content Creation (12.43\%) and Education \& Training (11.86\%), which aim to enhance human creativity and learning. For instance, by generating personalized video comments \cite{lin2024personalized} or by analyzing classroom dialogues to improve teaching quality \cite{ji2025classcomet}. More specialized domains like Security \& Surveillance (5.65\%) and Autonomous Driving (4.52\%) also leverage these technologies for critical tasks, including detecting abnormal events in security footage \cite{phan2025aadc} and predicting pedestrian intentions \cite{munir2025pedestrian}.

In contrast to these domain-specific efforts, a substantial portion of the literature pursues epistemic or infrastructural goals. Notably, Human-Robot/Agent Interaction appears as the single largest category (27.68\%). This line of work investigates the broader principles of human-machine interaction in everyday settings such as homes \cite{blanco2024ai}, public spaces \cite{jahangard2024jrdb}, or service environments \cite{xie2024empathetic}. These studies, rooted in human-computer interaction (HCI), emphasize the design and deployment of socially responsive agents and interactive systems, treating interaction itself as the primary object of inquiry. Similarly, 14.12\% of papers focus on advancing core social intelligence capabilities rather than specific applications. These researches develop generalizable models for basic social perception tasks, such as zero-shot action recognition \cite{dai2024gpt4ego} and human-object interaction detection \cite{gao2024contextual}. These foundational contributions provide the essential tools that researchers in the application-driven domains can then adapt and deploy.

\begin{figure}[h]
    \centering
    \includegraphics[width=.90\linewidth]{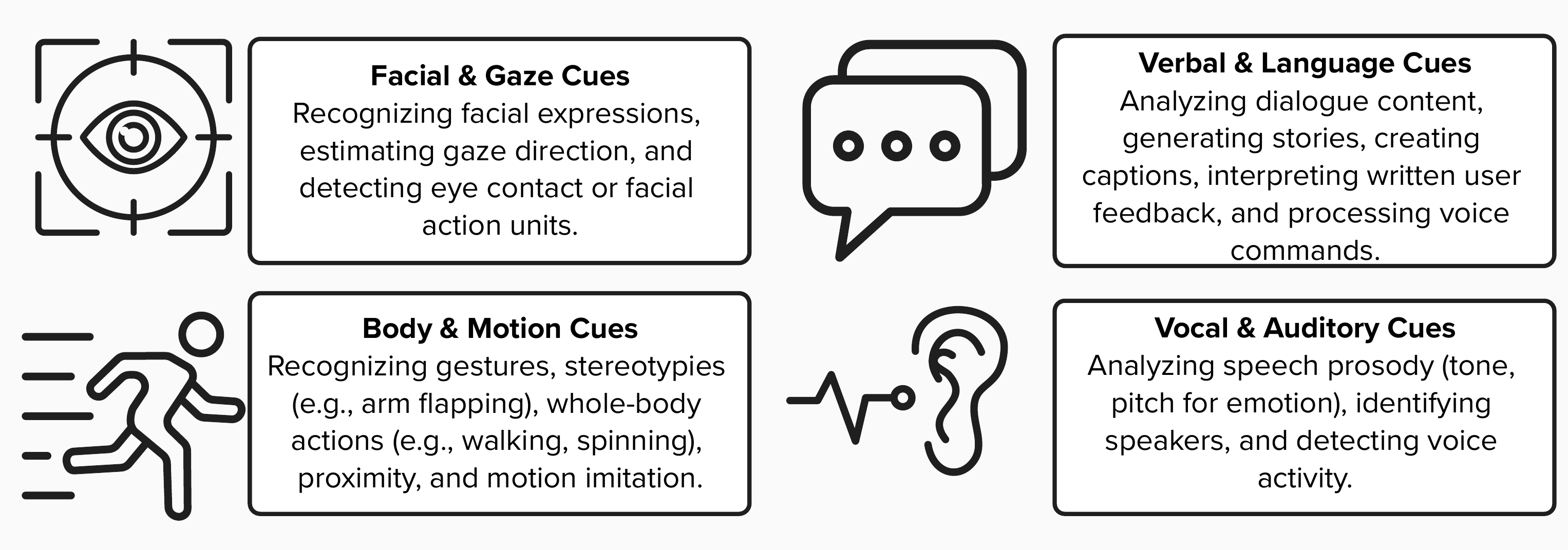}
    \caption{\textbf{Four Major Categories of Behavioral Cues Analyzed in Reviewed Systems.} The figure displays the primary types of human behavioral cues that the surveyed systems are designed to interpret. These include verbal and language cues, body and motion cues, vocal and auditory cues, and facial and gaze cues.}
    \label{fig:cue}
\end{figure}

We identified four major categories of behavior cues from our corpus (Figure. \ref{fig:cue}), reflecting the diverse ways systems interpret human activity. The most common are verbal and language cues (61.93\%), which include analyzing written or spoken dialogue, generating captions, and responding to voice commands. This is followed by body and motion cues (48.29\%), such as gesture recognition, action classification, and motion tracking. These two categories dominate the field, indicating a strong focus on linguistically and physically expressive forms of behavior. Other less commonly used behavior types include vocal and auditory cues (23.29\%), which involve speech prosody, speaker identification, or emotion detection through audio, as well as facial and gaze cues (22.16\%), including recognition of expressions or eye contact. Notably, these nuanced non-verbal cues, often essential in human social perception, remain relatively underutilized. 

\subsection{\textbf{Behavior Characteristics}}

\begin{figure}[h]
    \centering
    \includegraphics[width=0.7\linewidth]{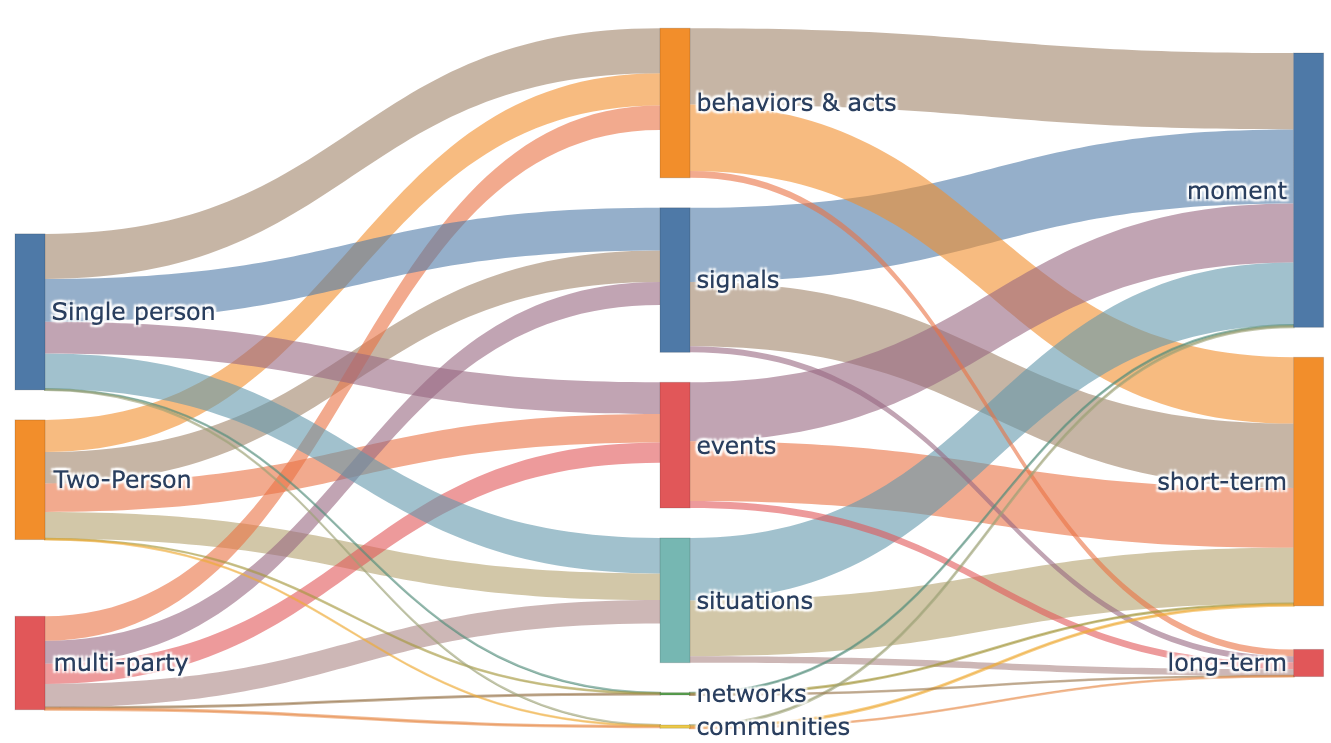}
    \caption{\textbf{Analysis of Social Behavior Along Three Key Dimensions.} This Sankey diagram visualizes the relationships between the interaction structure, social units, and temporal scale in the reviewed literature. The flow highlights a dominant research focus on analyzing individual behaviors ("single person") at a micro-level ("signals," "behaviors \& acts") over brief timescales ("moment," "short-term").}
    \label{fig:sankey}
\end{figure}

To characterize the nature of behaviors analyzed in LLM-powered multimodal systems, we examined three key dimensions: interaction structure (i.e., the number of interacting participants), temporal scale (i.e., the time span of behavior under analysis), and social units (i.e., the building blocks of social analysis). Together, these dimensions offer a structured view of the complexity of human behavior addressed by current systems (Figure. \ref{fig:sankey}).

A clear pattern emerges across the corpus: most studies center on micro-level, short-term, and individually framed behavior. The temporal scope of analysis is heavily skewed toward the immediate (94.3\%) and short-term (82.9\%), with only a small minority (8.5\%) addressing long-term behavior. That is, most systems are designed to interpret actions and interactions within a single session or isolated moment, rather than track evolving dynamics over days or weeks. 

Examination of social units further confirms this micro-level orientation: the dominant social units analyzed are behaviors and acts (98.3\%) and signals (95.4\%)—that is, discrete gestures, expressions, utterances, or cues such as gaze and tone. In contrast, more complex structures such as communities (43.2\%) or networks (2.8\%) are comparatively rare. Even in cases where papers touch on group-level constructs (e.g., “classroom community” or “team collaboration”), the analytical unit is often reduced to individual actions or attributes within that context. For instance, studies like \cite{abe2025classification} dissect speaker behaviors in group conversations based on fine-grained individual cues rather than modeling the collective dynamic. \cite{echeverria2025teamvision} and \cite{li2023hierarchical} offer notable accounts of network structures: the former analyzes collaboration in medical teams through sociograms and communication networks, while the latter constructs character networks in films through knowledge graphs.

This tendency toward “parallel individual analysis in group contexts” is further reflected in the interaction structures observed. Single-person analysis dominates (63.6\%), even when conducted within multi-party environments. Studies rarely analyze contingent interactions between multiple participants (32.9\% for two-person, 26.7\% for multi-party), and even fewer model relational dynamics such as turn-taking or coordination patterns.

\subsection{\textbf{System Characteristics}}

To complement our analysis of behavior characteristics, we next examine the context in which systems are deployed and interact, focusing on two key dimensions: embodiment degree and system role. Together with the behavior layer, these form what we refer to as a two-layered social context for socially intelligent systems. While behavior characteristics capture the nature of social inputs that systems interpret, system characteristics reflect how the system itself is situated and engaged within interactional settings.

The majority of reviewed papers ( 61.93\%) fall into the N/A category for embodiment degree, indicating a strong emphasis on non-agentic systems, that is, systems which analyze or generate behavior but do not take on an interactive social role. Among embodied systems, physically embodied agents (21.02\%) are more prevalent than virtual embodiments (8.52\% each for both embodied and disembodied forms), suggesting a notable interest in deploying social intelligence in physical, real-world environments (e.g., service robots, therapeutic companions \cite{xie2024empathetic, lima2025promoting}), rather than purely digital interfaces.

System role refers to the system’s functional position in the interaction,such as acting as a passive responder, an active participant, or an external observer. We observe that most systems (57.39\%) function as observers, reflecting the dominance of a post hoc analytical paradigm in which LLMs are used to interpret recorded social behavior. A smaller number of systems (22.73\%) operate as passive responders, such as in Q\&A or dialogue completion tasks. Notably, only 35 papers (19.89\%) describe systems that act as active participants, meaning systems capable of steering interaction, providing instructions, or setting goals (e.g. \cite{tanneberg2024help, ito2025robot, skantze2025applying}). 

Taken together, these findings reinforce a key asymmetry in current research: while many systems are designed to analyze complex social behaviors, relatively few are positioned as socially capable agents within those interactions. Bridging this gap—between observing social complexity and participating in it—remains a central challenge for the development of socially intelligent multimodal systems.

\subsection{\textbf{Operationalized Social Intelligence Competencies}}

Prior research in psychology has conceptualized social intelligence as comprising both cognitive and behavioral facets: the former involves interpreting others’ mental states and social norms, while the latter refers to acting appropriately in social situations \cite{weis2005social}. However, operationalizing these facets for AI systems remains an open challenge. Drawing on performance-based models in psychology \cite{weis2005social} and recent frameworks proposed for AI agents \cite{mathur2024advancing}, we adapted a six-part taxonomy of competencies as the foundation for our coding:

\begin{itemize}
    \item \textbf{Social perception.} The system’s capacity to perceive and discriminate socially relevant signals from multimodal inputs (e.g., emotion recognition, gaze tracking).
    \item \textbf{Social knowledge.} The ability to apply contextual or normative information to interpret social situations (e.g., social norms, roles, expectations).
    \item \textbf{Social memory.} The ability to retain and retrieve socially relevant information across time (e.g., remembering faces or past interactions).
    \item \textbf{Social reasoning.} The capacity to infer hidden states, intentions, or causal relationships from observed cues (e.g., intent or moral reasoning).
    \item \textbf{Social creativity.} The ability to produce contextually novel and socially effective outputs rather than replicating patterns..
    \item \textbf{Social interaction.} The ability to engage in contingent, co-regulated exchanges with humans or agents in real time, adapting dynamically to the interactional flow.
\end{itemize}

To examine how LLM-powered multimodal systems instantiate these competencies, we coded each system for the presence of the six capabilities. Our results reveal clear trends: social perception (100\%), social knowledge (97.74\%), and social Reasoning (95.48\%) are by far the most commonly operationalized competencies. This aligns with our earlier findings that most systems emphasize fine-grained recognition of expressive behavior (e.g., facial expressions, gestures, verbal cues), interpretive functions rooted in norms and context (e.g., classroom roles, clinical routines), and inference of hidden states (e.g., emotion, intent, collaboration dynamics). This foundational stack of Perception, Knowledge, and Reasoning forms the cognitive engine of nearly every system we reviewed.

Complementing this cognitive foundation is social memory (69.49\%), which supports more context-sensitive analysis by allowing systems to store and recall prior information, such as user preferences or conversational history. Far less common, however, are systems that implement social interaction (41.81\%) or social creativity (18.08\%). This gap underscores a broader asymmetry across our corpus: while systems are skilled at analyzing and interpreting human social behavior, far fewer are designed to participate in or co-construct social exchanges in real time. This is consistent with our earlier observation that most systems function as observers rather than actors, and that behavior modeling remains primarily individual and immediate rather than collective or longitudinal.

\begin{figure}[h]
    \centering
    \includegraphics[width=1\linewidth]{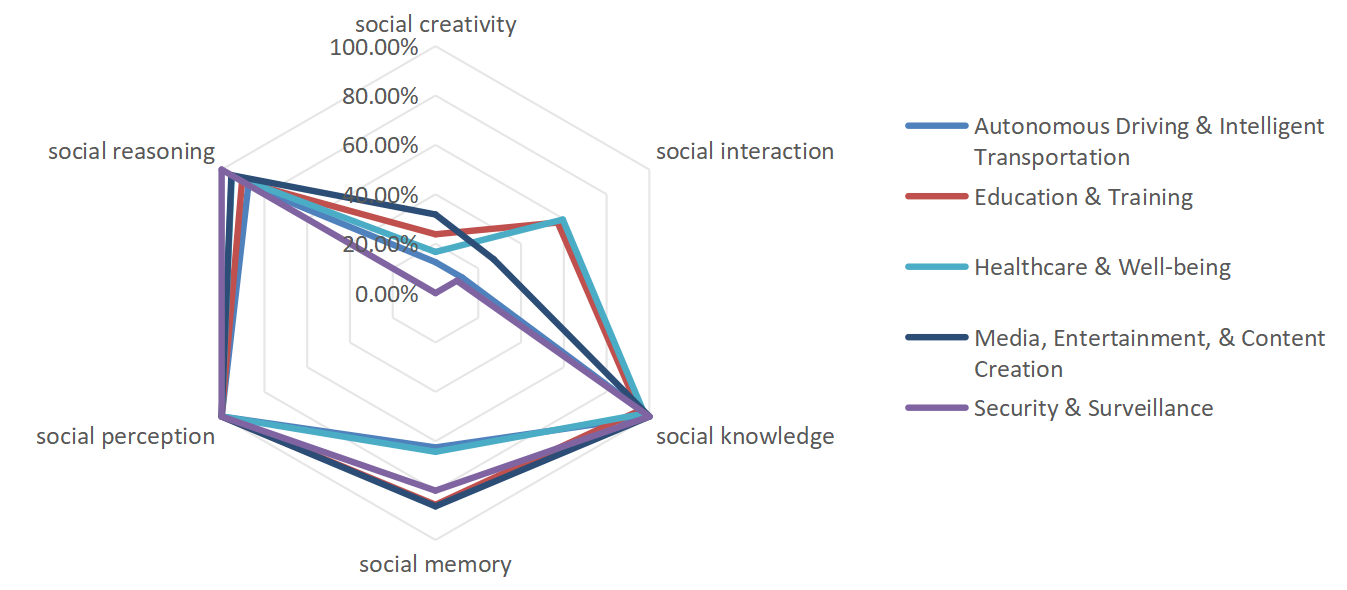}
    \caption{\textbf{Distribution of Social Intelligence Competencies Across Application Domains.} This radar chart compares the implementation of six key social intelligence competencies across five major application domains. The chart shows that competencies like social perception and reasoning are nearly universal, while social creativity and social interaction are far less common, particularly in domains like Security \& Surveillance.}
    \label{fig:radar}
\end{figure}

This asymmetry is also domain-dependent. As shown in Figure~\ref{fig:radar}, domains such as Healthcare and Well-being and Media \& Entertainment more frequently engage creative or interactional competencies—e.g., by generating empathetic responses or dynamic social content—while Security \& Surveillance systems are heavily concentrated around perception and reasoning.

\begin{samepage}
\begin{callout}
\textbf{Major Takeaways--Scope and Nature of Social Intelligence (RQ1)}
\begin{itemize}[leftmargin=*, nosep]
    \item \textbf{Dominance of Perception \& Recognition:} Nearly all of the 176 works focus on social perception (100\%) and reasoning (95\%), while social interaction (41.8\%) and creativity (18\%) are rarely implemented.
    \item \textbf{Short-Term, Individual-Centric Focus:} 94\% of studies analyze immediate or short-term behavior and 63\% focus on single-person analysis, with very limited modeling of long-term trajectories or multi-party dynamics.
    \item \textbf{Uneven Cue Usage:} Verbal \& language cues dominate, while nuanced non-verbal signals like gaze, facial expressions, and prosody are underutilized despite their importance for social intelligence.
    \item \textbf{Observer Role Prevails:} Most systems act as post-hoc observers rather than active participants, highlighting a gap between analyzing and co-constructing social interaction.
\end{itemize}
\end{callout}
\end{samepage}

%% file: section/4Results-rq2.tex
\section{RQ2: Technical Operationalization of LLM-powered Multimodal Systems in Human Behavior Understanding}
\label{results:rq2}

To address our second research question, we examined how social intelligence is technically implemented across LLM-powered multimodal systems. Our analysis focused on three key dimensions: (1) how input modalities are handled and transformed, (2) what modalities are processed by the core LLM, and (3) what operationalization strategies are employed—i.e., whether through architectural changes, prompting, or model fine-tuning. 

\subsection{\textbf{Modality Reflection}}

\begin{figure}[h]
    \centering
    \includegraphics[width=1\linewidth]{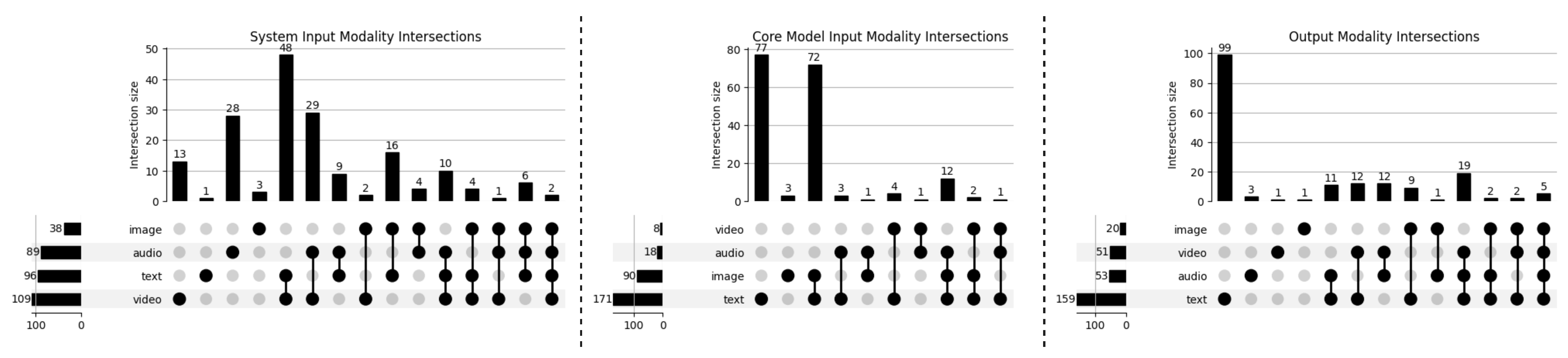}
    \caption{\textbf{The Modality-to-Text Bottleneck: A Visual Journey from System Input to Final Output.} This series of charts visualizes the transformation of data modalities as they are processed by the reviewed systems, revealing a distinct "modality-to-text" pipeline. \textbf{System Input Modality Intersections (Left)}: This chart shows that systems are designed to capture rich, multimodal data. The most frequent combination involves video, audio, and text together (48 papers), indicating an ambition to analyze complex, real-world social interactions. \textbf{Core Model Input Modality Intersections (Center)}: This chart illustrates a critical shift in the pipeline. Rich inputs like video and audio are rarely processed directly by the core LLM. Instead, they are overwhelmingly converted into text (used alone in 77 papers) or a combination of text and static images (72 papers), simplifying dynamic signals for language-centric reasoning. \textbf{Output Modality Intersections (Right)}: This chart shows the final output of the systems. The process culminates in a predominantly unimodal, text-only result (99 papers), reinforcing the role of these systems as analytical engines that distill a complex multimodal reality into a linguistic summary.}
    \label{fig:modality_combined}
\end{figure}

Across the reviewed papers, a staged and text-centric processing architecture emerges as the dominant design pattern. Systems typically begin with rich multimodal inputs—video, audio, and text—capturing the complexity of human social behavior. As shown in Figure \ref{fig:modality_combined} (left), video (109 papers), audio (89 papers), and text (96 papers) are the most common input sources. The most frequent combination is video + audio (48 papers), reflecting a growing interest in audiovisual social interaction, such as analyzing online conversations \cite{wang2024semeval}.

However, a significant discrepancy appears when comparing system input modalities with the actual inputs passed into the core LLM (Figure \ref{fig:modality_combined} (center)). While system-level pipelines often begin with raw video or audio data, these signals are rarely processed directly by the LLM. Instead, they are transformed into text (e.g., transcripts, captions) or static images through upstream modules like speech recognition, visual labeling, or temporal sampling. Text is overwhelmingly the dominant core input modality, appearing in 171 out of 176 papers.

These conversions highlight a design bias toward simplifying temporally rich and socially expressive modalities into more LLM-compatible formats. While this alignment enables efficient reasoning through LLMs, it may also introduce limitations—potentially compressing nuanced social signals such as gaze dynamics, conversational timing, or vocal affect into overly coarse representations.

Finally, this technical pipeline culminates in a predominantly uni-modal, text-based output. As shown in Figure \ref{fig:modality_combined} (right), text is the primary form of system output, with text-only output appearing in 99 papers, far exceeding any other modality or combination. This reinforces our earlier finding of systems as "analytical engines": the complex processing of a rich, multimodal social reality is ultimately distilled into a linguistic output, such as generating stories for cognitive therapy \cite{blanco2024ai} or providing writing assistance to users \cite{cai2024pandalens}.

\subsection{\textbf{Operationalization Technique}}

To further analyze how social intelligence is implemented, we categorized each system based on three technical approaches:
\begin{itemize}
    \item Architecture-centric: Changes to system-level structures or perception pipelines.
    \item Inference-centric: Use of prompting or other runtime configurations.
    \item Model-centric: Fine-tuning or modifying the LLM weights.
\end{itemize}

Architecture-centric techniques dominated the field, appearing in 164 papers. These typically involved custom pipelines for multimodal perception or modules to scaffold interaction. Inference-centric approaches, used in 137 papers, relied on prompt engineering or conversational role-playing to adapt behavior at runtime. Model-centric techniques, such as fine-tuning or adapter training, were the least common (82 papers), likely due to higher technical complexity.

Hybrid strategies were prevalent. Over 40\% of the papers used both architecture- and inference-level strategies  (e.g. \cite{bravo2024improving}), and 52 papers employed all three methods simultaneously (e.g. \cite{chakhtouna2024multi}). These full-stack approaches often paired prompting with tailored pipelines and lightweight model adaptation.

Across all strategies, system design consistently centered around a narrow set of large language models. Over half of the papers used GPT-family models, including GPT-3.5, GPT-4, GPT-4V, and GPT-4o, as well as related variants. While other LLMs such as Claude, LLaVA, Vicuna, Mistral, and Gemini were present, their use was comparatively sparse. Notably, few papers used Gemini models despite their multimodal capabilities \cite{mardanbegi2025gazelog}, and some papers did not report their base LLM at all \cite{dangol2025want}. This reliance on GPT-based architectures suggests a lack of diversity in model experimentation.

\begin{samepage}
\begin{callout}
\textbf{Major Takeaways--Technical Operationalization (RQ2)}
\begin{itemize}[leftmargin=*, nosep]

    \item \textbf{Modality-to-Text Bottleneck:} Multimodal inputs (video/audio) are typically converted to text transcripts or keyframes before reaching the LLM, stripping away timing, prosody, and subtle context cues.
    \item \textbf{Text-Dominant Outputs:} Over half of the systems output text-only results, reinforcing their role as analytical engines rather than multimodal interactive agents.
    \item \textbf{Architecture-Centric Designs Dominate:} 93\% rely on custom pipelines for preprocessing; model fine-tuning and inference-level adaptation are less common, treating LLMs as mostly fixed reasoning cores.
    \item \textbf{Model Homogeneity:} GPT-family models are used in the majority of studies; few explore alternatives like Gemini, LLaVA, or open-source models, limiting insights on model-specific performance trade-offs.

\end{itemize}
\end{callout}
\end{samepage}

%% file: section/4Results-rq3.tex
\section{RQ3: Evaluation Methodologies and Metrics for LLM-powered Multimodal Systems in Human Behavior Understanding}

\subsection{\textbf{Evaluating Social Intelligence: a divided field}}
\label{sec:rq3_1}
To address our third research question, we analyzed the methods and metrics used in the literature to evaluate socially intelligent MLLMs. The analysis reveals a significant divergence in evaluation practices, characterized by a dominant, machine-centric paradigm focused on technical correctness and a less common yet more diverse, human-centric paradigm focused on human impact.

Out of 176 papers, 10~(5.6\%) did not conduct any form of evaluation. Among those that did, 89~(50.5\%) conducted only system evaluations using custom or established benchmarks and common ML metrics (e.g., accuracy); 4~(2.2\%) conducted only human evaluations (e.g., interviews/surveys) including qualitative feedback metrics such as enjoyment, comprehension; and 73~(41.4\%) conducted both system and human evaluations.

A common pattern is to report \emph{Classification/Recognition Accuracy} as the premier metric for system evaluation, i.e., success measured by correspondence to ground truth (e.g., mAP in~\cite{madan2025mip}). Authors also report \emph{Computational Efficiency} (e.g., FPS in~\cite{bravo2024improving}) and \emph{Language Generation Quality} (e.g., BLEU in~\cite{ziaeetabar2024hierarchical}) to capture purely technical performance. This aligns with our earlier finding that many systems are designed as “analytical engines”.

Human evaluations adopt a more holistic approach. The most common methods are \emph{Surveys} (34.6\%) and \emph{Interviews} (19.3\%). Reported human-centered metrics span three aspects:
\begin{enumerate}
    \item Usability \& User Experience: assessing if the system is "good to use" through measures like the System Usability Scale (SUS)~\cite{lima2025promoting};
    \item Task Performance \& Behavioral Change: measuring if the system is "effective" through metrics like learning gain (pre-post test scores~\cite{liu2024classmeta});
    \item Perception of AI Qualities: evaluating if the system is "likeable" or "trustworthy" by assessing factors such as perceived robot empathy (RoPE scale)~\cite{xie2024empathetic}.
\end{enumerate}
This indicates that when researchers adopt a human-centric perspective, they emphasize a technology’s value and impact in authentic human contexts.

\subsection{\textbf{System Evaluation Mechanism}} 

To bridge the gap between HCI goals and AI tasks in evaluating socially intelligent LLM systems, we further investigated how these systems were evaluated in terms of \emph{System Evaluation} and compiled Table~\ref{tab:bench_mark_table}. The table provides a comprehensive list of benchmarks and the associated papers in which they were used to evaluate socially intelligent multimodal LLM systems. It is organized by Application Area (Human Goal) and Computational Task (Model Framing) to guide future researchers in selecting appropriate benchmarks and prior evaluation techniques for their analysis goals.

For papers with system evaluation, 101 papers used benchmarks (55 established, 17 mixed, 29 new). These benchmarks cover four \textbf{Application Areas} related to human goals when analyzing social multimodal data: (1) \emph{Action \& Activity Understanding}, (2) \emph{Social \& Affective Intelligence}, (3) \emph{Video \& Language Comprehension}, and (4) \emph{Domain-Specific Applications}. These goals map to common \textbf{Computational Tasks} in machine learning (e.g., 1.1 Classification \& Localization, 3.1 VideoQA \& Reasoning) and to more nuanced doemain specific tasks (e.g., 2.1 Social Reasoning \& Theory of Mind (ToM), 4.1 Healthcare \& Accessibility) that are benchmark-evaluable.

Tasks are then mapped to \textbf{Evaluation Resources}—benchmarks/datasets the authors either (a) used as established (e.g., RefCOCO~\cite{gao2024contextual}, GazeFollow~\cite{gupta2024exploring}), (b) derived/built upon (e.g. Playlogue\cite{kalanadhabhatta2024playlogue}, OSCaR\cite{nguyen2024oscar}), or (c) created as custom resources for nuanced human behavior analysis (e.g., SMILE~\cite{hyun2023smile}, Autism Restricted and Repetitive Behavior Dataset (ARRBD)\cite{wang2025ms}). 

\begin{samepage}
\begin{callout}
\textbf{Major Takeaways--Evaluation Practices (RQ3) }
\begin{itemize}[leftmargin=*, nosep]
  \item {\textbf{Machine-Centric Benchmarking Dominates:}  Across 176 papers, 10 (5.6\%) had no evaluation; of the rest, 89 (50.5\%) were system-only, 4 (2.2\%) human-only, and 73 (41.4\%) both; human methods centered on Surveys (34.6\%) and Interviews (19.3\%).}
  \item {\textbf{Benchmark Uses and Updates:} Researchers both rely on traditional benchmarks and increasingly adapt or introduce new ones to evaluate nuanced multimodal social tasks: among 162 papers with system evaluation, 101 (62.3\%) used benchmarks—55 (54.5\%) established, 17 (16.8\%) adapted/derived, and 29 (28.7\%) newly created.}
  \item {\textbf{Dataset Directory--Mapping Goals→Tasks→Resources:} (see Table~\ref{tab:bench_mark_table}). The table compiles benchmarks and their associated papers, organized by Application Area (Human Goals) and Computational Tasks (Model Framing), providing researchers a practical guide to select appropriate benchmarks and system evaluation techniques for their specific analysis goals.}
\end{itemize}
\end{callout}
\end{samepage}

\label{app:bench} 
\renewcommand{\arraystretch}{1.5} 
\begin{longtable}{>{\raggedright\arraybackslash}p{0.17\linewidth}|>{\raggedright\arraybackslash}p{0.20\linewidth}|>{\raggedright\arraybackslash}p{0.55\linewidth}}
    \caption{\textbf{Bridging HCI goals and AI tasks in evaluating socially intelligent LLM systems.} Entries are organized by Application Area (Human Goal), Computational Task (Model Framing), and Evaluation Resource (Benchmark/Dataset). Dataset names include citations from papers they've been used in.}
    \label{tab:bench_mark_table} \\
    \toprule
    \midrule
        \textbf{Application Area 
        (Human~Goal)} & 
        \textbf{Computational Task (Model~Framing)} & 
        \textbf{Evaluation Resource (Benchmark~/~Dataset)}
        \\
        \midrule
        \midrule
    \endfirsthead 

    \multicolumn{3}{c}%
    {{\bfseries \tablename\ \thetable{} -- continued from previous page}} \\
    \toprule
    \midrule
        \textbf{Application Area 
        (Human~Goal)} & 
        \textbf{Computational Task (Model~Framing)} & 
        \textbf{Evaluation Resource (Benchmark~/~Dataset)}
        \\
        \midrule
        \midrule
    \endhead 

    \midrule
    \bottomrule
    \endlastfoot 
        Action \& Activity Understanding &
        Anomaly \& Safety Monitoring &
        AN-Workout\cite{phan2025aadc};
        UCF-Crime\cite{phan2025aadc};
        XD-Violence\cite{phan2025aadc}.
        \\
        &
        Anticipation \& Forecasting &
        \textbf{Gaze:}
        AVA\cite{gupta2024exploring};
        ChildPlay\cite{gupta2024exploring};
        EMS\cite{madan2025mip};
        ENCAA\cite{madan2025mip};
        GazeFollow\cite{gupta2024exploring};
        MIP-GAF[\cite{madan2025mip};
        MS\cite{madan2025mip};
        NCAA\cite{madan2025mip}.
        \\
        &
        Classification \& Localization &
        \textbf{A/V:}
        AVS benchmark\cite{gao2024contextual};
        COCO-20i\cite{gao2024contextual};
        PASCAL-5i\cite{gao2024contextual};
        RefCOCO(+/g)\cite{gao2024contextual}
        \textbf{|}          
        \textbf{Ego:}
        CharadesEgo\cite{dai2024gpt4ego};
        Charades\cite{lin2023match};
        EGTEA\cite{dai2024gpt4ego};
        EPIC-KITCHENS-100\cite{dai2024gpt4ego,pasca2023summarize};
        OSCaR(derived from EPIC-KITCHENS)\cite{nguyen2024oscar}
        Ego-Exo4D\cite{seino2025expert};
        Ego4D \cite{liu2024human,nguyen2024oscar, wang2025less,rodin2024action,mohaiminul2024video,he2025designminds,pasca2023summarize};
        EgoSchema\cite{mohaiminul2024video}
        \textbf{|}
        HMDB51\cite{chen2025vision,lin2023match,wei2025svmfn};
        K600\cite{lin2023match};
        Kinetics\cite{chen2025vision,wei2025svmfn};
        MiT\cite{lin2023match};
        MiniSSv2\cite{lin2023match};
        NTU RGB+D(60, 120)\cite{chen2025vision,chen2024vision};        
        PKU-MMD\cite{chen2024vision};
        UAG-(FunQA, SSBD, OOPS\cite{abdullah2025ual}) ;
        UAV\cite{lin2023match};
        UCF101\cite{chen2025vision,lin2023match,wei2025svmfn};
        ``Chaotic World'' benchmark\cite{chen2024leveraging};
        Charades-STA\cite{abdullah2025ual};
        Subsets of Charades(ToM)\cite{rodriguez2025integrating}.
        \\
        &
        Interaction, Pose \& Motion &
        3DPW\cite{feng2025posellava};
        AGD20K\cite{wang2025affordstruct};
        HICO-(DET\cite{gao2024contextual,ren2024learning},IIF\cite{wang2025affordstruct});
        Human3.6M\cite{feng2025posellava};
        HumanML3D(ToM)\cite{zhou2024avatargpt};
        PoseFix\cite{feng2025posellava};
        PosePart\cite{feng2025posellava};
        PoseScript\cite{feng2025posellava};
        Reasoning-based Pose Estimation(RPE)\cite{feng2024chatpose};
        SWIG-HOI\cite{gao2024contextual};
        Speculative Pose Generation(SPG)\cite{feng2024chatpose};
        V-COCO\cite{ren2024learning};
        \\
        &
        Procedures \& State Change &
        50 Salads dataset\cite{wang2025multimodal};
        Breakfast dataset\cite{wang2025multimodal};
        COIN\cite{patel2023pretrained};
        CrossTask\cite{patel2023pretrained};
        Annotated version of the GTEA dataset(gaze, ToM)\cite{matsukawa2024data};
        KIT Bimanual Action\cite{ziaeetabar2024hierarchical};
        MS-COCO\cite{ziaeetabar2024hierarchical};
        Multiple Object States Transition (MOST) dataset\cite{tateno2025learning};
        VirtualHome simulator[151];
        Youcook2\cite{ziaeetabar2024hierarchical};
        Annotated dataset derived from YouCook2(ToM)\cite{hori2025interactive};
        \\
        & 
        Retrieval \& Long-Form Understanding &
        ACM MM 2023 Grand Challenge\cite{li2023hierarchical};
        NIST TRECVID 2022 DVU\cite{li2023hierarchical};
        RSL dataset\cite{ding2024integrating};
        UVSD dataset\cite{ding2024integrating};
        Waldo\cite{alper2023learning} (retrieval);
        Wenda\cite{alper2023learning} (retrieval);
        subset of imSitu\cite{alper2023learning} (retrieval)
        \\\midrule
        Social \& Affective Intelligence&
        Dialogue \& Interaction & 
        Playlogue\cite{kalanadhabhatta2024playlogue}
        \\
        &
        Emotion \& Affect &
        AffectNet\cite{rajesh2025enhancement};
        CASME2\cite{lu2024gpt};
        CelebA-(base, dialog)\cite{lu2024face};
        DEEP corpus\cite{muller2024recognizing};
        DISFA\cite{lu2024gpt};
        ECF2.0(sourced from 'Friends')\cite{wang2024semeval};
        EMOTIC\cite{teotia2024evaluating,etesam2024contextual}(Subset of EMOTIC(ToM)\cite{yang2023contextual});
        E³ dataset\cite{feng20243}
        GoEmotions\cite{rajesh2025enhancement};
        Hume-Vidmimic2\cite{qiu2024language};
        IEMOCAP\cite{chakhtouna2024multi,feng2024foundation};
        MELD\cite{feng2024foundation};
        MSP-(Improv, Podcast)\cite{feng2024foundation};
        RAF-DB\cite{lu2024gpt,lu2024face};
        Real-Life Trial dataset\cite{lu2024gpt};
        SCB-dataset3\cite{teotia2024evaluating};
        iMiGUE\cite{lu2024gpt}
        \\
        &
        Social Reasoning \& Theory of Mind (ToM) &
        CH-SIMS\cite{liu2024multimodal};
        CK+\cite{liu2024multimodal};
        CMU-MOSI\cite{liu2024multimodal};
        Friends\cite{liu2024funnynet};
        Interpersonal Relation Dataset (IPR)\cite{akay2025interpersonal};
        JAFFE\cite{liu2024multimodal};
        KDEF\cite{liu2024multimodal};
        LMU Munich Executive Leadership Perception (LMU-ELP\cite{martin2024larger};
        MHD\cite{liu2024funnynet};
        MMLSCU dataset\cite{meng2024mmlscu};
        MMTOM-QA(ToM)\cite{jin2024mmtom};
        MUSTARD\cite{liu2024funnynet};
        Memento10k\cite{martin2025parameter};
        Modified version of Learning to Listen (L2L) dataset\cite{ng2023can};
        MuMA-ToM\cite{shi2025muma};
        MultimodAl Tamil Hate (MATH) dataset\cite{mohan2025multimodal};
        People in Social Context (PISC)\cite{akay2025interpersonal};
        SMILE\cite{hyun2023smile};
        Social-IQ 2.0\cite{agrawal2024listen};
        Subset of 'Automobile Cabin Voice Interaction Data' dataset(ToM)\cite{moeini2024calibrated};
        TBBT\cite{liu2024funnynet};
        UR-Funny\cite{liu2024funnynet};
        USC's Split-Steal corpus (newly annotated)\cite{han2024knowledge};
        VCC2018(ToM)\cite{basit2024tinydigiclones};
        VVALUES\cite{wang2025multimodal};
        VideoConviction\cite{galarnyk2025videoconviction};
        VCC2018(ToM)\cite{basit2024tinydigiclones}
         \\\midrule
        Video \& Language Comprehension &
        Captioning \& Narration &
        Activity-Net\cite{dong2024m2};
        DiDemo\cite{dong2024m2};
        Flickr8K\cite{thayaparan2023digital};
        LSMDC\cite{dong2024m2,han2023autoad};
        MAD-eval\cite{han2023autoad};
        MSR-VTT\cite{adnan2024irag, dong2024m2};
        StreetAware\cite{adnan2024irag};
        Tokyo MODI\cite{adnan2024irag};
        VATEX\cite{dong2024m2};
        VQA-v2\cite{adnan2024irag};
        Flickr8Knew(ToM)\cite{thayaparan2023digital}
        \\
        &
        Retrieval \& Grounding &
        ActivityNet-Captions\cite{xu2024vtg};
        Charades-STA\cite{xu2024vtg};
        QVHighlights\cite{xu2024vtg};
        QueryYD dataset\cite{ning2024spica}
        \\
        &
        Semantic Graphs \& Story &
        MovieGraphs+\cite{dong2025key,qin2025scenario};
        PerVidCom\cite{lin2024personalized};
        ViSR+\cite{qin2025scenario}
        \\
        &
        VideoQA \& Reasoning &
        AGQA 2.0\cite{wang2024efficient};
        ActivityNet-QA\cite{ding2025language};
        EgoSchema\cite{wang2024efficient};
        IntentQA dataset\cite{li2023intentqa};
        Kinetics\cite{ding2025language};
        MSRVTT-QA\cite{ding2025language};
        MSVD-QA\cite{ding2025language};
        NExT-(GQA, QA)\cite{wang2024efficient};
        QA-Ego4D\cite{shen2024encode};
        Something-Something V2\cite{ding2025language};
        Video-MME\cite{ding2025language} 
        \\\midrule
        Domain-Specific Applications &
        Driving \& Scene Understanding &
        \textbf{Driving Actions:}
        BDD100K(ToM)\cite{xiao2024chatcam};
        CIFAR-10/100(ToM)\cite{xiao2024chatcam};
        DHPR (Driving Hazard Prediction)\cite{charoenpitaks2024exploring};
        HighD\cite{manzour2024rag};
        JAAD\cite{munir2025pedestrian,manzour2024rag};
        Derived from KITTI\cite{jain2024semantic};
        NuScenes\cite{jain2024semantic};
        PIE\cite{munir2025pedestrian};
        PSI\cite{manzour2024rag};
        PedPrompt\cite{munir2025pedestrian};
        \textbf{Street Scene Understanding:}
        StreetAware\cite{arefeen2024trafficlens,arefeen2024vita};
        Tokyo MODI\cite{arefeen2024vita}
         \\
         &
         Education &
         Kyoto University's lectures\cite{ito2025robot};
         Coursera educational videos\cite{ji2025classcomet};
         YouTube educational videos\cite{ji2025classcomet};
         \\
         &
         Healthcare \& Accessibility &
         \textbf{ASD \& Autism:} 
         Audio-Visual Autism Spectrum Dataset (AV-ASD)\cite{deng2024hear};
         Autism Restricted\cite{wang2025ms};
         Repetitive Behavior Dataset (ARRBD)\cite{wang2025ms};
         \textbf{Accessibility:}
         Derived from VideoA11y-40K dataset\cite{li2025videoa11y}
         \textbf{Sign Language Recognition:} 
         AUTSL\cite{liu2025signpepper};
         Isolated Sign Language Recognition Corpus(ToM)\cite{lim2024enhancing}; MSASL\cite{liu2025signpepper};
         WLASL100\cite{liu2025signpepper};
         \textbf{General:}~MEDIQA-QS\cite{wen2024leveraging};
         MeQSum\cite{wen2024leveraging};
         Reddit posts\cite{wen2024leveraging}
        \\
         &
         Human-Robot Interaction &
         DataHRC\cite{li2024toward};
         Disfluent Navigational Instruction Audio Dataset(DNIA)\cite{sun2024trustnavgpt};
         HHIRChat\cite{shakti2024ov};
         JRDB-Social\cite{jahangard2024jrdb};
         RoboCup@Home GPSR task benchmark\cite{shirasaka2024self};
         RoboTHOR\cite{sun2024trustnavgpt};
         \\
\end{longtable}



%% file: section/4Results-rq4.tex
\section{RQ4: Ethical Challenges and Risks of LLM-powered Multimodal Systems in Human Behavior Understanding}

We distinguish two outcomes for each ethics issue reported in a paper: \textit{Addressed Ethical Risks}—risks explicitly acknowledged \emph{and} accompanied by a concrete control—and \textit{Remaining Ethical Risks}—risks explicitly raised as unresolved, emerging, or out-of-scope. We coded all ethics-relevant statements with the AIR’24 framework~\cite{zeng2024air}, which contains 314 categories in a four-tier taxonomy; at Tier~1 the families are \textit{System \& Operational}, \textit{Content Safety}, \textit{Societal}, and \textit{Legal \& Rights}. In parallel, we coded papers for \textit{Implemented Mitigations} (actions or design choices actually used in the reported system/study) versus \textit{Proposed Mitigations} (unimplemented strategies suggested for future work or deployment).

\subsection{\textbf{Coverage of Ethical Considerations (Paper-Level)}}
\label{results:rq4_2}
\begin{figure}[h]
    \centering
    \includegraphics[width=0.75\linewidth]{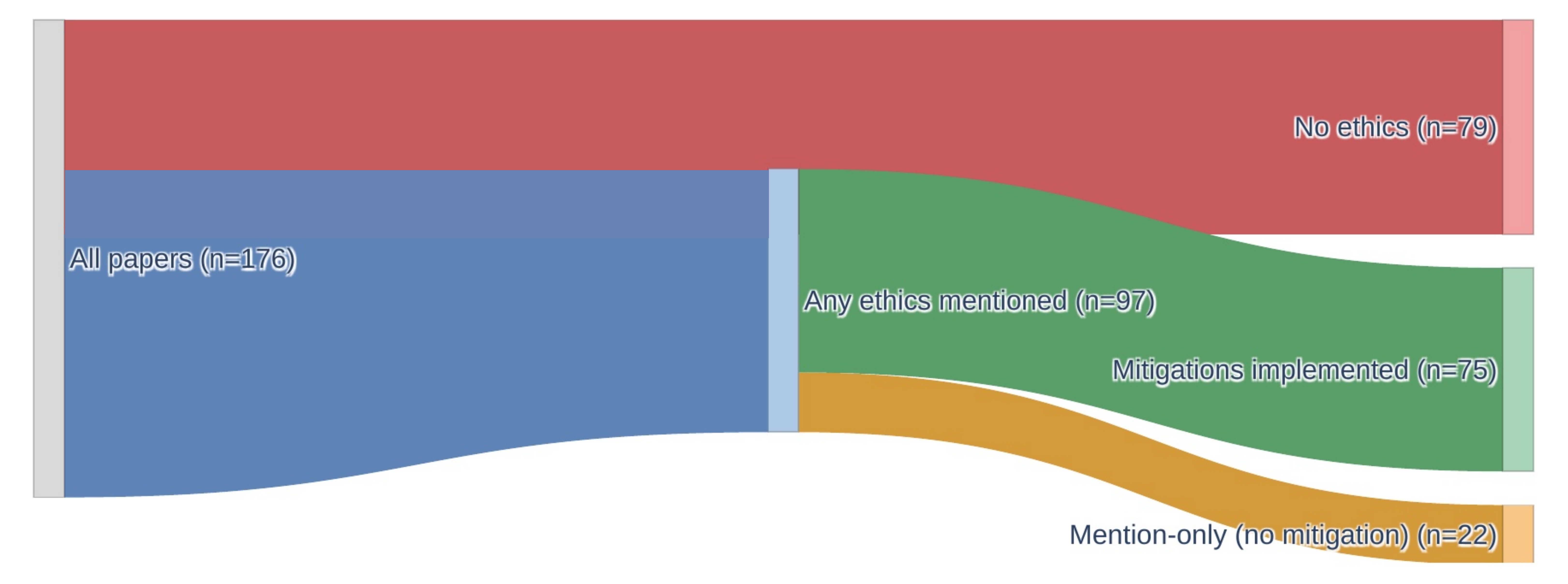}
    \caption{\textbf{Paper-Level Coverage of Ethical Considerations.} This Sankey diagram shows the breakdown of ethical engagement across the 176 reviewed papers. It reveals that 45\% of papers do not mention ethics. Of the 55\% that do, 77\% report implementing at least one mitigation, while 23\% only mention risks without describing a concrete action.}
    \label{fig:sankey_ethics_overall}
\end{figure}

Figure~\ref{fig:sankey_ethics_overall} summarizes paper-level coverage as a funnel: of \textbf{176} papers, \textbf{97} (\textbf{55\%}) mention ethics in any form, while \textbf{79} (\textbf{45\%}) do not. Among the ethics-aware set, \textbf{75/97} (\textbf{77\%}) report at least one \emph{implemented} mitigation, whereas \textbf{22/97} (\textbf{23\%}) \emph{only} mention risks without action. In absolute terms, this means \textbf{43\%} of all papers (75/176) include any concrete mitigation, and \textbf{12.5\%} (22/176) stop at mention-only.

This analysis reveal that ethical practice remains unevenly integrated into LLM-powered multimodal work. Even among ethics-aware papers, \emph{about a quarter} stop at acknowledgement without intervention, and \emph{nearly half} of the literature omits ethics entirely. In the following subsubsections, we examine \emph{which} categories receive mitigation and \emph{how} they are addressed, proceeding through the AIR’24 taxonomy: \textbf{Tier~1} (families), \textbf{Tier~2} (subfamilies), and \textbf{Tier~3} (subcategories). This structure surfaces where interventions are reported and where gaps remain without presupposing outcomes.

\subsection{\textbf{Risk Families: What Gets Addressed vs.\ What Remains Open}}
\begin{figure}[h]
    \centering
    \includegraphics[width=1\linewidth]{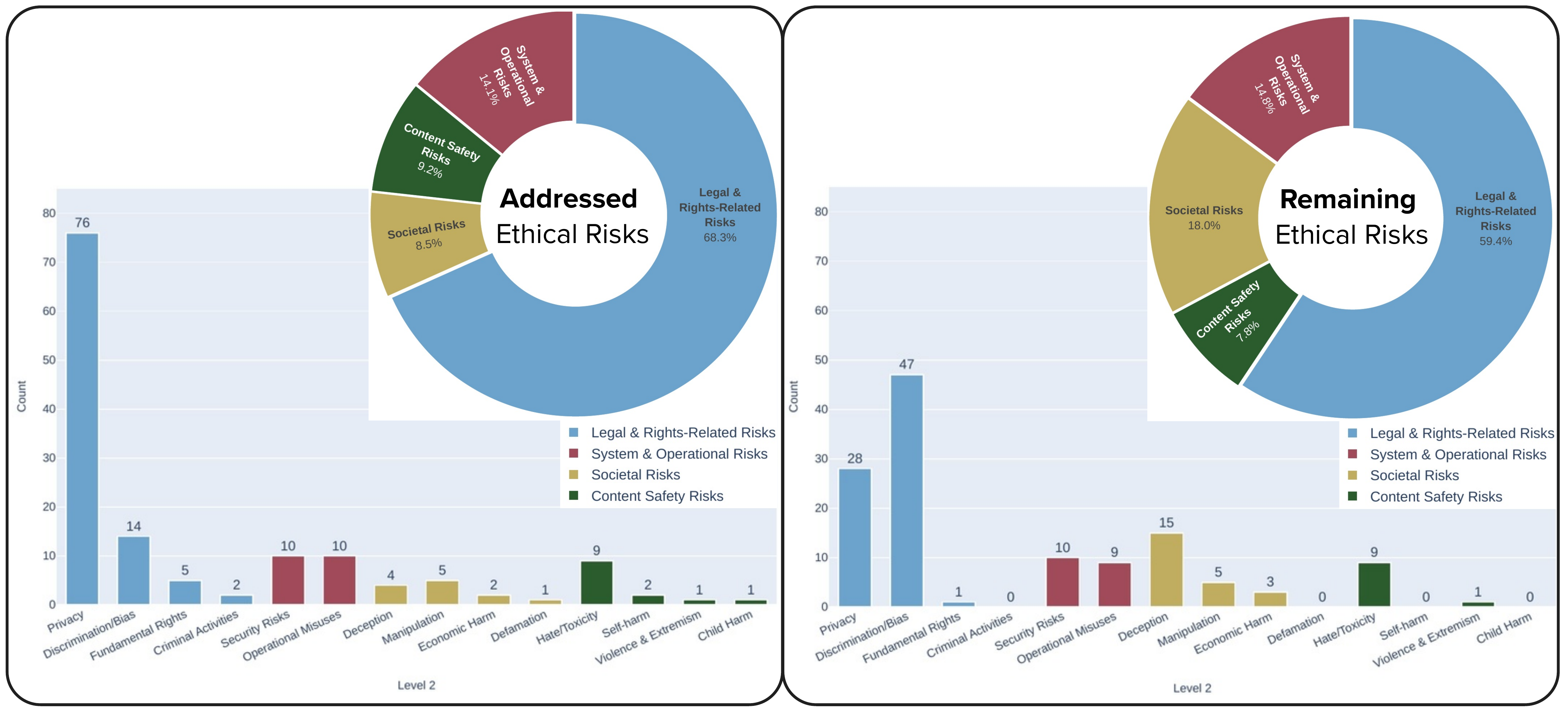}
    \caption{\textbf{Comparison of Addressed vs. Remaining Ethical Risk Mentions.} This figure compares the types of ethical risks that are actively addressed with mitigations versus those that are mentioned as remaining or open challenges. The charts highlight a clear shift: addressed risks are dominated by concerns over Legal \& Rights, particularly Privacy. In contrast, remaining risks show a significant increase in concerns related to Discrimination/Bias and broader Societal harms like deception.}
    \label{fig:ethics_comparison}
\end{figure}

Figure~\ref{fig:ethics_comparison} (Tier~1 and 2) and our full coding (Tier~3 below) point to a consistent pattern: {Legal \& Rights} dominates what gets \emph{addressed} (97 mentions), while {Societal} concerns are more visible among what \emph{remains} (23 vs.\ 12). {System \& Operational} is comparatively stable across buckets (20 vs.\ 19), and {Content Safety} stays smaller overall (13 vs.\ 7).

Within {Legal \& Rights}, the emphasis flips: \emph{addressed} items center on \textit{Privacy} (76) with far fewer \textit{Discrimination/Bias} mentions (14), whereas \emph{remaining} items are led by \textit{Discrimination/Bias} (48) over \textit{Privacy} (28). {System \& Operational} splits similarly in both buckets (\textit{Security Risks} 10 and \textit{Operational Misuses} 10 addressed; 10 and 9 remaining). For {Societal}, \textit{Deception} grows sharply among remaining (17 vs.\ 4 addressed), while \textit{Manipulation} appears in both (5 addressed; 3 remaining). {Content Safety} is broader in the addressed set (\textit{Hate/Toxicity} 9 plus \textit{Violence \& Extremism}, \textit{Child Harm}, \textit{Self-harm}), but thinner among remaining (mostly \textit{Hate/Toxicity} 6 plus \textit{Violence \& Extremism} 1).

Privacy-focused subcategories dominate addressed mentions (\textit{Unauthorized Privacy Violations} 46; \textit{Types of Sensitive Data} 30) but are much smaller among remaining (19 and 9). By contrast, fairness-related items cluster in the remaining set (\textit{Discriminatory Activities} 26; \textit{Protected Characteristics} 22 vs.\ 8 and 6 addressed). {Security} is symmetric (\textit{Confidentiality} 6; \textit{Integrity} 4 in both). {Operational Misuses} diverge: \textit{Autonomous Unsafe Operation} (6 addressed vs.\ 2 remaining) and \textit{Advice in Regulated Domains} (4 addressed; 5 remaining), with \textit{Automated Decision-Making} appearing only among remaining (2). Societal harms show a similar skew: \textit{Mis/disinformation} is higher among remaining (13 vs.\ 4 addressed), while \textit{Misrepresentation} appears in both (5 addressed; 3 remaining). Addressed {Content Safety} spans more severe categories (\textit{Self-harm} 2; \textit{Child Harm} 1) than the remaining set, which concentrates on toxicity variants (e.g., \textit{Hate Speech}, \textit{Offensive Language}, \textit{Harassment}) at lower counts.

The field reports concrete controls where pathways are established—especially \textit{privacy/security}—but fairness and socially grounded harms surface more strongly as \emph{remaining} issues. Figure~\ref{fig:ethics_comparison} (Tier~1 and 2) outlines this shift, and the Tier~3 details above show where, specifically, privacy is operationalized (e.g., unauthorized disclosure; sensitive data types) while fairness (\textit{discriminatory activities}, \textit{protected characteristics}) and deception (\textit{mis/disinformation}) persist. Complete results table for these 3 tier coding is provided in Appendix.

\subsection{\textbf{Mitigations in Practice vs.\ On the Horizon}}
Implemented mitigations cluster around what papers can control immediately in deployment: (1) \emph{privacy-by-design} (on-device/edge or local models; de-identification, masking/blurring, restricted releases; secure storage/configuration and no-retention settings; IRB/consent, opt-out flows, pseudonyms, and controlled lab studies); (2) \emph{human oversight and gating} (therapists/caregivers or instructors in the loop; researcher moderation; explicit audience transparency about virtual agents; post-hoc identity disclosure in studies; professor control over classroom robots); (3) \emph{guardrails on generation and execution} (content filters and guideline-constrained prompting; verbosity caps; pre-generated outputs to avoid non-determinism; RAG/ReAct/template scaffolds; sandboxed interpreters, input validation, and “predefined functions only” for code execution); (4) \emph{embodied/interactive safety engineering} (motion-planning constraints, simulation-before-execution, scene-graph anomaly detection, intent-aware planning; model-side explainability and recovery affordances such as inner speech, tiered summaries, or KG+fuzzy-rule rationales); (5) \emph{data/process governance} (use of pre-approved or de-identified datasets; licensing/GDPR-aware collection; restricted sharing; cloud configurations disabling retention/training); and (6) \emph{user controls and protective UX} (moment-selection to filter captured content; color-blind modes; supportive default content to displace toxic inputs; expert-editable behavior codebooks).

Proposed (unimplemented) directions extend beyond immediate controls toward harder open problems: (1) \emph{fairness and generalization} (dataset diversification, bias audits, culturally sensitive fine-tuning; replication across tasks/populations; annotator diversity, informed codebooks, stereotype-aware modeling); (2) \emph{technical privacy and data control} (differential privacy, homomorphic encryption, modular disabling of PII processing, generative PII replacement; silent-speech input; bystander indicators/recording lights and local face-blurring; machine unlearning; on-device-by-default for future deployments); (3) \emph{transparency, verification, and evaluation} (interpretable outputs with citations; trust cues and confidence scores; standardized evaluation APIs; human verification/deferral for sensitive judgments; user-driven fact checking; deployment guidelines, audits, and accountability channels; offline fallbacks); (4) \emph{robust autonomy under uncertainty} (uncertainty handling/denoising; voice-activity detection for turn-taking; temporal reasoning; multi-sensor/video expansion; improved trajectory planning, feedback-based replanning, memory and intention modeling); (5) \emph{person-centered UX} (customization/personalization of agents and haptics/visual parameters; tapered intervention to reduce over-reliance; accessibility features; designs that amplify existing human bonds); and (6) \emph{computational pragmatics and safety-of-content} (model distillation and energy-aware operation; stronger filters with human-in-the-loop moderation; clearer annotation/wording to avoid refusals). In short, what is \emph{implemented} centers on privacy, consent, and controllability in today’s pipelines; what is \emph{proposed} targets fairness, reliable reasoning under uncertainty, and accountable, socially situated behavior.

\begin{samepage}
\begin{callout}
\textbf{Major Takeaways--Ethical Challenges and Risks (RQ4)}
\begin{itemize}[leftmargin=*, nosep]
  \item {\textbf{Ethics is Unevenly Addressed:} 45\% of reviewed papers do not mention ethics, and 23\% of those that do stop at acknowledgment without implementing mitigation. 
  This gap is visible in the funnel (Figure~\ref{fig:sankey_ethics_overall}).}
  \item {\textbf{What gets fixed vs. what lingers.} Reported mitigations cluster in \emph{privacy/security} (e.g., data handling, consent), while \emph{fairness} and broader \emph{social harms} (mis/disinformation, manipulation) surface more often as remaining issues. (Figure~\ref{fig:ethics_comparison}).}
  \item {\textbf{Near-term controls vs. hard problems.} Implemented actions emphasize controllability—on-device processing, de-identification, human oversight, sandboxed execution, and prompt/RAG scaffolds—whereas proposed strategies target the harder frontiers: fairness/generalization, transparent verification, and uncertainty-aware autonomy in socially situated settings.}
\end{itemize}
\end{callout}
\end{samepage}

%% file: section/5Discussion.tex
\subsection{Beyond Cognitive Recognition: Toward Behavioral and Interactive Social Intelligence (RQ1)}

Our coding reveals a consistent imbalance across competencies: almost all papers target social perception, social knowledge, and social reasoning, with many also using social memory, while social interaction and social creativity remain far less common. Systems are built to detect, recall, and reason about social cues, yet far fewer attempt to act on them in contextually appropriate, adaptive ways. In short, the systems we reviewed largely optimize the cognitive facet of social intelligence but underserve the behavioral facet, precisely the duality emphasized in psychology and in recent AI-oriented syntheses\cite{mathur2024advancing, goleman2006social}. 


Several patterns in system design help explain this imbalance. Most systems adopt an observer role, analyzing social signals post-hoc rather than participating actively in interaction. They also tend to focus on immediate, short-term behavior (94\% of studies) and largely consider individuals in isolation, leaving multi-party and longitudinal dynamics underexplored
(63\%). Furthermore, verbal and language cues dominate system inputs, while nuanced non-verbal signals such as gaze, facial expressions, and prosody are rarely utilized, limiting the richness and adaptability of social behaviors these systems can generate.

\subsubsection{Social Knowledge remains generic, not norm-sensitive.} 

In psychology, social knowledge is understood as a complex competency encompassing not only declarative facts about the social world but also procedural knowledge of social norms and etiquette that guide behavior in specific contexts \cite{conzelmann2013new, weis2005social}. Crucially, this knowledge is not universal; it is highly dependent on cultural values and situational specifics.

Our review, however, finds that in the LLM-based systems we surveyed, this competency is frequently operationalized as generic, text-based commonsense rather than as situated, norm-sensitive knowledge. For example, IntentQA \cite{li2023intentqa} leverages an LLM to provide commonsense context for disambiguating intent, such as knowing that "putting a spoon in one's mouth" likely means "eat food." While useful for reasoning about individual actions, this approach overlooks the procedural knowledge needed to navigate norms in larger communities \cite{mathur2024advancing}. 

A notable exception is ClassMeta \cite{liu2024classmeta}, where a virtual agent embodies the social knowledge specific to a classroom "community." The agent's behaviors, such as knowing when to break the silence or how to correct off-topic chatter, are forms of procedural knowledge specific to that learning environment. The scarcity of such systems highlights a significant gap, and the emergence of benchmarks like EgoNormia \cite{rezaei2025egonormia}, designed to evaluate understanding of physical social norms, signals a growing recognition that AI must move beyond generic commonsense toward more nuanced, norm-sensitive social knowledge.

\subsubsection{Cognitive Modeling Choices Shape the Behavioral and Interactive Gap}

The way social knowledge and creativity are modeled in current systems amplifies the behavioral gap \cite{mathur2024advancing}. Many systems are designed to converge on a single “correct” output rather than generate multiple possibilities or flexibly adapt \cite{mori2025comprehensive}. This convergent problem-solving approach is misaligned with the divergent nature of social creativity and interaction, which often require exploring multiple behavioral options and adapting to dynamic social contexts \cite{raz2024open}. Combined with still-developing capabilities, systems seemed to follow a reactive pattern: they map recognized inputs to predefined responses rather than co-constructing interactions in real time.

The short-term, individual focus of most research further limits opportunities to model multi-party coordination or longitudinal interaction trajectories. Meanwhile, the heavy reliance on verbal cues, with non-verbal signals largely underutilized, restricts sensitivity to the subtle information that humans use to coordinate social behavior. Together, these factors foster a reactive, observer-style design: systems can respond to detected inputs but are not structured to engage adaptively or collaboratively.

Advancing the interactive dimension of social intelligence therefore requires a deliberate shift. Future systems must move beyond perception and reasoning to support agents that can act creatively, adaptively, and socially in complex, real-world contexts. Only by integrating norm-sensitive social knowledge, multi-party and longitudinal modeling, and richer multimodal cue processing can AI systems begin to bridge the cognitive-behavioral gap and participate meaningfully in social interactions. 

\subsection{Modality Reflection: Building Social AI Beyond Textual Bottlenecks (RQ2)}

Across the corpus, “multimodal” pipelines frequently compress rich signals into simpler, language-ready surrogates: video becomes frames or captions; audio becomes transcripts. This choice appears to be a pragmatic adaptation to the task at hand, for many of the specific goals outlined in these papers, a complete, native understanding of the original multimodal signal is not required. The simplified data is often sufficient for their descriptive task. However, this compression inevitably leads to a loss of nuance, creating a critical gap between what the system can process and how humans communicate. 

This methodological constraint leads to systemic limitations. For example, preprocessing pipelines that convert video to keyframes and transcripts for content coding are effective at identifying concrete elements but demonstrate reduced accuracy for abstract concepts like emotion or communication style, as they exclude the very modalities through which that content is conveyed \cite{liu2024harnessing}. Similarly, rendering personal media searchable by converting inputs into text-based annotations and embeddings inherently discards non-textual information. This conversion process results in the loss of small-scale visual details (e.g., brand logos) and eliminates the contextual associations between discrete events (e.g., notebook imagery and meeting contexts) \cite{li2025omniquery}. 

These examples illustrate a fundamental architectural constraint: current multimodal systems prioritize computational tractability over communicative fidelity. The resulting frameworks capture explicit content ("what was said" or "what appears in a frame") while operating independently of the social, temporal, and contextual frameworks through which human communication derives meaning.

We also observe a pattern of visual and textual dominance and an under-use of the audio channel in our reviewed papers. While vision encoders are widely used, audio is often reduced to its transcript or, in more sophisticated cases, represented by simplified surrogates. This approach is exemplified by Hyun et al. \cite{hyun2023smile}, who converted prosody (e.g., pitch, intensity) into numerical features within a text prompt to help LLMs reason about laughter. Conversely, Ng et al. \cite{ng2023can} took this to its logical extreme by excluding audio entirely, successfully generating listener reactions from transcripts alone. This text-only success suggests why audio may be de-prioritized: transcripts appear "good enough"—preserving semantic and temporal cues while avoiding the complexity of audio processing.

Yet, the limitations of this text-only approach are defined by the authors themselves. Ng et al. \cite{ng2023can} report that the model is limited in cases where socially diagnostic cues are audible but not in the text, such as sarcastic jokes where vocal tone contradicts the literal words. This underscores the importance of the audio channel, a point powerfully reinforced by the findings of Hyun et al. \cite{hyun2023smile} The fact that their simplified numerical surrogates for prosody significantly improved performance over text-alone is compelling evidence of the information density within the original audio signal: if even the "ghost" of the signal is this impactful, the signal itself must be vital.

These design choices reflect a common architectural constrain: multi-modal inputs are converted to textual representations to enable LLM processing. This pattern suggests an underlying assumption that symbolic linguistic representations can adequately capture multimodal social communication. However, recent empirical work challenges this "logocentric" assumption \cite{westmoreland2022multimodality}. Xu et al. \cite{xu2024academically} document only a "low correlation" between linguistic and social intelligence in LLMs, observing "superficially friendly" styles without contextual grounding. This aligns with fundamental questions in social AI about whether language can serve as a sufficient intermediate representation for social signals that may not be "effectively described in language" \cite{mathur2024advancing}. This critique parallels broader calls to "align perception with language models" \cite{huang2023language}, reflected in their assertion that "Language Is Not All You Need." 

\subsection{Optimized for Benchmarks, Misaligned with Human Context: A Call for Ethically-Informed Social Evaluation (RQ3-RQ4)} 

Our review reveals a critical disconnect between how socially intelligent multimodal systems are developed and the real-world contexts they are meant to inhabit. The predominant evaluation paradigm skews heavily toward machine-centric benchmarking over human-centered assessment. Our analysis of 176 papers shows that 50.5\% report system-only evaluations, while 41.4\% evaluate on both, and a mere 2.2\% rely on human evaluation alone (see~\S\ref{sec:rq3_1}). This focus risks optimizing for performance on narrow, quantitative benchmarks that often fail to predict success in the messy, dynamic reality of human social interaction. Such systems may learn to exploit dataset-specific heuristics—a phenomenon known as "shortcut learning"—rather than developing robust, generalizable social understanding~\cite{geirhos2020shortcut, gururangan2018annotation}. Even as benchmarks evolve from simple recognition to more complex reasoning tasks, they continue to privilege structured, static outcomes, overlooking the reciprocal, adaptive qualities that define genuine social intelligence~\cite{eriksson2025trust, ibrahim2024towards}. Consequently, benchmark progress becomes a poor proxy for meaningful social impact~\cite{eriksson2025trust}.

This evaluation gap is compounded by an ethics gap. Our findings show that ethical considerations are addressed unevenly: while a slight majority of papers (55\%) mention ethics, fewer than half (43\%) implement any concrete mitigation(see~\S\ref{results:rq4_2}). Furthermore, existing mitigations tend to cluster around technically tractable issues like privacy and data security. Systemic risks related to fairness, discrimination, and broader societal harms—the very issues that manifest most acutely upon deployment—remain largely unaddressed, relegated to the category of "future work."

Taken together, these patterns create a significant deployment gap. Current research primarily reports on proxy indicators of success, such as benchmark scores and technical efficiency, rather than the human outcomes that matter in practice. As foundational HCI scholarship reminds us, true impact is measured by tracing the path from system inputs and outputs to real-world outcomes, accounting for the complex context in which they occur~\cite{goodman2012observing}. Without socially-grounded evaluation and concrete ethical controls, the development of multimodal LLMs remains optimized for publication, not for people.

Ignoring this twofold gap creates a direct pathway from benchmark success to real-world harm. Models that appear socially competent in controlled tests can be brittle in practice. This is supported by empirical work revealing a low correlation between the linguistic polish of LLMs and their actual social intelligence, often resulting in a "superficially friendly" style that is ungrounded in context~\cite{sap2022neural}. This brittleness is amplified by evaluation practices that reward success on simplified tasks or biased datasets, as was seen with early benchmarks like Social-IQ, where models could exploit statistical patterns rather than engaging in genuine social reasoning~\cite{guo2023desiq}.
When deployed in sensitive domains, the failure modes are predictable and severe: 1) In mental healthcare and related public service contexts, misreading non-verbal cues by emotion recognition systems can lead to dangerous, inappropriate clinical responses and exacerbate existing inequities or harms~\cite{roemmich2024emotion}. 2) In education or counseling, where an embodied agent’s synthetic rapport—an empathetic tone, a nodding avatar—lends unearned credibility to incorrect guidance~\cite{steenstra2025risk}. 3) In hiring and lending, where biased analysis of a candidate's facial expressions or speech patterns from video interviews can perpetuate and scale discrimination~\cite{raghavan2020mitigating}. In short, benchmark-optimized and ethically under-engineered systems do not just underperform; they externalize social and legal risks onto the very users and communities they are intended to serve. These are precisely the risks—fairness, bias, and societal harm—that our analysis shows are persistently under-mitigated in current multimodal systems. 

To bridge this deployment gap, we advocate for a pragmatic re-orientation that tightly couples evaluation with ethics. We propose that researchers \emph{evaluate for the behaviors they intend to deploy and co-measure success with risk exposure in the same study.} This involves several concrete steps:

\textbf{1) Integrate Metrics:} Pair quantitative system metrics with qualitative, human-in-the-loop evaluations that assess interaction quality, including reciprocity, responsiveness, transparency, and trust~\cite{amershi2014power, dove2017ux}. \textbf{2) Benchmark for Doing, Not Naming:} Design evaluation tasks that require wise or appropriate action in realistic scenarios (e.g., de-escalating a conflict, providing an empathetic response), not just the labeling of static data. \textbf{3) Treat Ethics as Implementation:} Move beyond acknowledgement. Make at least one concrete mitigation a standard for any ethics-aware study—whether through privacy-by-design, human-in-the-loop governance, or robust content guardrails—with focused attention on the fairness and societal harms where mitigation currently lags. \textbf{4) Report for Impact, Not Proxies:} Frame results in terms of the full input → output → outcome chain, interpreting performance alongside measured risk. 

Adopting this ethically-informed approach will help ensure that gains on benchmarks translate into credible evidence of socially competent, interaction-aware, and deployment-ready multimodal systems.

%% file: section/5Discussion-whole.tex
\subsection{Social Creativity: a Blue Ocean Opportunity for Designing LLM-Supported Social Intelligence} 


Our review reveals a landscape shaped by striking imbalances in how multimodal, LLM-powered systems are designed, deployed, and evaluated for social intelligence. While systems excel at perception and recognition, they fall short in enabling interaction and remain especially underdeveloped in fostering creativity. These systemic gaps do more than mark limitations—they point to a blue ocean opportunity for designing LLM-based multimodal solutions that not only interpret social behavior but also support and promote social creativity.

Social creativity emerges as the most underdeveloped competency in our review, reflecting a fundamental mismatch between its divergent nature and the convergent design paradigm of current AI systems. In psychology, social creativity is defined as the capacity to generate multiple, diverse interpretations or solutions for a given social situation, grounded in counterfactual reasoning and Theory of Mind (ToM) \cite{conzelmann2013new, weis2005social, mathur2024advancing}. Yet most systems are optimized to reduce ambiguity and converge on a single, most probable output. For example, intent recognition tasks that resolve uncertainty into one correct inference \cite{li2023intentqa}. This pursuit of ground truth is inherently at odds with the exploratory and generative character of creativity.

The root of this limitation lies in the weak foundation of what has been termed Neural Theory-of-Mind (N-ToM) in Large Language Models \cite{sap2022neural}. Although robust ToM is considered essential for modeling others’ mental states \cite{williams2022supporting}, empirical studies show that LLMs rely on shallow heuristics and spurious correlations rather than genuine reasoning \cite{sap2022neural, shapira2023clever}. Their apparent ToM quickly collapses under adversarial testing. The implication is clear: if systems cannot reliably perform convergent ToM tasks, they are even less equipped to engage in the divergent reasoning needed for creativity.

These limitations are evident in real-world applications. Some systems show early signs of creativity, such as generating novel gestures for humanoid robots \cite{huang2025emotion}, but outputs typically collapse into a single “optimal” behavior. By contrast, human-centered contexts—from creative tasks seeking alignment with a subjective “vibe” \cite{hammad2025s} to home-based speech therapy requiring adaptable strategies \cite{dangol2025want}—demand repertoires of possibilities rather than singular solutions. Without the capacity for divergent, flexible responses, AI systems remain analytical observers rather than co-creative partners. 

Beyond competency imbalances, the pursuit of social creativity also reveals structural gaps in modality, evaluation, and ethics as we identified in our earlier discussion. Divergent generation depends on nuanced multimodal signals (prosody, gaze, timing, embodied gestures) that are often stripped away when inputs are reduced to text or static frames \cite{mathur2024advancing, li2025perception}, raising doubts about whether textual abstraction alone can approximate human social nuance \cite{feng2024far}. Evaluation practices further constrain progress: prevailing benchmarks reward convergence on a single correct answer, privileging recognition accuracy over generative flexibility \cite{fu2024mme}. Metrics for “naming” thus fail to capture success in “doing,” where the goal is to generate multiple plausible pathways toward social outcomes \cite{mathur2024advancing}. Finally, creativity introduces distinct ethical stakes. While divergent generation can expand user agency, it also risks producing misleading or manipulative behaviors and overwhelming users with unbounded options \cite{mathur2024advancing}. Ethical considerations must therefore shift from post-hoc safeguards to design preconditions to ensure outputs enhance trust and agency.

Looking forward, these gaps point toward a research agenda centered on social creativity, which we conceptualize not as artistic innovation, but as flexibility in social interpretation. This requires systems that can leverage ToM and counterfactual reasoning to generate diverse interpretations for a single social situation (e.g., a student's silence) rather than converging on one label. The implications of achieving this are profound: it fosters a more human-like, contextually-sensitive AI that moves beyond labeling "engagement low" to suggesting possibilities ("student may be thinking, or hesitant"). Most importantly, this capability enables a paradigm shift from AI as a judge to AI as a partner, facilitating "AI-human co-interpretation." By providing an "interpretive space" instead of a singular answer, the system empowers users (e.g., therapists or teachers) to make more empathetic and ethically-grounded decisions. Social creativity thus represents the frontier for transforming AI from an analytical observer into a co-creative partner that helps humans explore and navigate complex social relations.

\subsection{Limitations and Future Work}
Our review provides a clear, rigorously gathered snapshot of multimodal LLM work on social signals while making our scope choices explicit. The corpus draws on major indexes within a defined time window (so some gray/venue-specific work may be absent); the systems we synthesize mainly process visual and auditory inputs—the primary channels for face-to-face social cues; and we focus on input multimodality (how models interpret rich human behaviors) rather than output generation, which may underplay biosensor-centric or generative strands; our keyword emphasis could favor audiovisual pipelines; and an LLM-assisted coding workflow can misclassify. We mitigated these by searching multiple databases, tailoring Boolean queries, full-text screening, documenting criteria, anchoring on a manually coded seed set, human-verifying all machine labels, and checking inter-rater reliability. Field-wide tendencies toward short, benchmark-centric evaluations bound generalizability; we treat these as directions, not deficits. Looking ahead, we will apply the proposed framework in more diverse contexts—and within these transparent boundaries, our synthesis remains trustworthy and actionable.